\documentclass[aip,jcp, preprint,graphicx]{revtex4-1}

\usepackage{xspace}
\usepackage{bm}
\usepackage{amsmath}
\usepackage[dvipsnames]{xcolor}
\usepackage[version=4]{mhchem}
\usepackage{physics}
\usepackage{siunitx}
\usepackage{hyperref}

\newcommand*{\duo}{\textsc{Duo}\xspace}
\newcommand*{\eg}{\textit{e.g.}\xspace}
\newcommand*{\etal}{\textit{et al.}\xspace}

\newcommand*{\ie}{\textit{i.e.}\xspace}
\newcommand*{\viz}{\textit{viz.}\xspace}
\newcommand*{\opH}{\mathcal{H}}
\newcommand{\iu}{\mathrm{i}\mkern1mu}
\newcommand*{\transT}{\mathsf{T}}
\newcommand*{\abinitio}{\textit{ab~initio}\xspace}
\newcommand{\exocross}{\textsc{ExoCross}\xspace}
\newcommand*{\VOXSigma}{\mathrm{X}\,^4\Sigma^-}
\newcommand*{\VOdouSigmap}{\mathrm{1}\,^2\Sigma^+}
\newcommand*{\VOAPi}{\mathrm{A}\,^4\Pi}
\newcommand*{\VOApPhi}{\mathrm{A}'\,^4\Phi}

\newcommand*{\red}[1]{{\color{red} #1}}

\newcommand*{\bmS}{\bm{S}}

\newcommand*{\bmI}{\bm{I}}

\draft

\begin{document}

\title{A variational model for the hyperfine resolved spectrum of VO in its ground electronic state} 

\author{Qianwei Qu, Sergei N. Yurchenko and Jonathan Tennyson}
\email[]{j.tennyson@ucl.ac.uk}
\affiliation{Department of Physics and Astronomy, University College London, London WC1E 6BT, United Kingdom}

\date{\today}
\red{The manuscript has been accepted by
The Journal of Chemical Physics}

\begin{abstract}
A variational model for the infra-red spectrum of VO is presented which aims to accurately predict 
the hyperfine structure within the VO $\mathrm{X}\,^4\Sigma^-$ electronic ground state.
To give the correct electron spin splitting of the $\mathrm{X}\,^4\Sigma^-$ state,
electron spin dipolar interaction within the ground state and 
the spin-orbit coupling between $\mathrm{X}\,^4\Sigma^-$ and 
two excited states, $\mathrm{A}\,^4\Pi$ and $\mathrm{1}\,^2\Sigma^+$, are
calculated \abinitio alongside hyperfine interaction terms.
Four hyperfine coupling terms are explicitly considered:
Fermi-contact interaction,
electron spin-nuclear spin dipolar interaction,
nuclear spin-rotation interaction and
nuclear electric quadrupole interaction. These terms are included as part of a full
variational solution of the nuclear-motion Schr\"odinger equation performed using 
program \textsc{Duo}, which is used to 
generate both hyperfine-resolved energy levels and spectra.
To improve the accuracy of the model, \abinitio 
curves are subject to small shifts.
The energy levels generated by this model
show good agreement with the recently derived empirical term values.
This and other comparisons
validate both our model and the recently developed hyperfine modules in \textsc{Duo}.
\end{abstract}


\maketitle

\section{Introduction}
Vanadium monoxide (VO) is an open shell diatomic molecule which absorbs strongly in the near infrared and visible
region of the spectrum. 
These absorptions are  of importance for astrophysics
where VO is known to be an important component of the atmosphere of cool stars. \cite{09Bernath.VO}
Recently attention has turned to the possible role of VO in the atmospheres of exoplanets
where it has been suggested that alongside TiO, VO absorption can change the temperature
profile of the planet's atmosphere.\cite{10MaSexx.VO}
Some tentative detections
of VO in exoplanet atmospheres have been reported \cite{16EvSiWa.VO,17TuLeBi.VO,17PaChPr.VO,jt699,20GoMaDrSi.VO,20LeWaMa.VO}
but none of these can be regarded as secure. 
There are two reasons for this. 
First, the spectra
of VO and TiO are heavily overlapped making them very hard to disentangle
at low resolution. 
Secondly, while the availability of a high-resolution
TiO line list suitable for high-resolution spectroscopic studies \cite{jt672} has led
to the confirmation of TiO in exoplanetary atmospheres, \cite{17NuKaHa.TiO,21SeSnMo.TiO,22BiHoKi.TiO}
the corresponding VO line list \cite{jt644} is not of sufficient accuracy
to be used in similar studies.\cite{22DeKeSn.VO}
Both the TiO and VO line lists cited
were produced using similar methodology by the ExoMol project \cite{jt810} but
a major difference between them is due to the underlying atomic physics.
While $^{16}$O and $^{40}$Ti both have nuclear spin, $I$, equal to zero,
the dominant isotope of vanadium, 
\ce{^{51}V}, has $I = 7/2$. 
The  interaction between the spin of unpaired electrons
and the nuclear spin
yields a very pronounced hyperfine structure which manifests
itself at even moderate resolution. This hyperfine
structure reduces parts of the \ce{^{51}V^{16}O} spectra to 
``blurred chaos
at Doppler-limited resolution'' \cite{89Merer.VO}.
Progress in identifying VO in exoplanetary atmospheres
using high resolution spectroscopy requires the development
of a model which includes a treatment of these hyperfine
effects. 
These effects were not considered in the ExoMol VOMYT line list.\cite{jt644}

A full survey of available high resolution spectroscopic data for VO
has recently been completed by Bowesman \etal \cite{jt869} as part
of a MARVEL (measured active rotation vibration energy levels) study of the system.
The nuclear hyperfine structure of \ce{^{51}V^{16}O}
has been measured \cite{81HoMeMi,82ChHaMe,91SuFrLoGi,95AdBaBeBo,08FiZixx.VO}
and modeled by effective Hamiltonians.\cite{94ChHaHu.VO,08FiZixx.VO}
However, for the the $\VOXSigma$ ground electronic state,
the experiments only gave the 
hyperfine constants for the lowest  ($v=0$)
vibrational level and therefore
provide limited information for
the observations of hot VO spectra involving higher vibrational levels.

Hyperfine structure in molecular spectra are usually
treated using perturbation-theory based 
effective Hamiltonians; these are usually accurate enough to
reconstruct the energy levels using the assumption that
hyperfine effects arise from  small perturbations. Thus, effective
Hamiltonians are 
widely used for fitting measured hyperfine-resolved 
energies or transitions, see Refs.\onlinecite{94ChHaHu.VO,08FiZixx.VO}
for examples involving VO.
However, the VOMYT line list \cite{jt644} shows that
interactions 
between the electronic states
reshape the line positions and intensities of VO.
Although we focus on the $\VOXSigma$ electronic ground state of VO in this paper,
the spin-orbit couplings between the low-lying $\VOXSigma$ and $\VOdouSigmap$
states as well as the  $\VOXSigma$ and $\VOAPi$ states are also
included in our model with the aim of obtaining the
correct spin splittings for the $\VOXSigma$ state.
This allows us to construct a full, predictive spectroscopic model of the  ground state which can be used as
input to the variational,  diatomic spectroscopic program \duo \cite{jt609} which we have recently extended to give
a full variational treatment of hyperfine effects.\cite{jt855}
This paper presents the development of this model.

\section{Computational details}

The electronic structure of VO has been 
investigated in numerous works.
\cite{95BaMaxx.VO,00BrRoxx.VO, 01CaSiAn.VO,01BrBoxx.VO, 03DaDeYa.VO,03Pyvaxx.VO,10KuMaxx.VO,07MiMaxx.VO,15HuHoHi.VO,jt623,21JiChBo.VO}
Among them,
the results for excited states represented by multi-reference configuration interaction (MRCI) wavefunctions
are more accurate.
\cite{07MiMaxx.VO,15HuHoHi.VO,jt623,21JiChBo.VO}
The most recent one by McKemmish \etal \cite{jt623}
laid the basis of the ExoMol VO linelist, VOMYT.\cite{jt644}
We also perform MRCI level calculations in this work
to get the potential energy curves (PECs) and spin-orbit coupling curves
for the electronic states of interest.
The electron spin-dipolar interaction and
hyperfine coupling curves of $\VOXSigma$
were obtained at the complete active space self consistent field (CASSCF) level.

\subsection{Quartet states}
In this work,
the potential energy and 
spin-orbit coupling curves
are calculated using MOLPRO 2015 \cite{MOLPRO2015}
at the MRCI level.
The energies are also improved by adding a
Davidson correction (+Q).

First, the ground $\VOXSigma$ state was calculated on its own
to avoid effects from other electronic states.
The active space used is larger than employed by McKemmish {\it et al.},
\cite{jt623}
as the work of 
Miliordos \etal \cite{07MiMaxx.VO}
shows that
the occupation of 4p orbitals of vanadium is not negligible.
In this work,
the 1s orbital of oxygen and
the 1s, 2s, 2p, 3s, 3p orbitals 
of vanadium were treated as
doubly occupied. 
The active space 
includes the
2s, 2p orbitals of oxygen
and 4s, 3d, 4p orbitals of vanadium.
In the four irreducible representations of $C_{2v}$ group, 
\viz $a_1, b_1, b_2, a_1$,
the numbers of occupied orbitals
are $(12, 5, 5, 1)$ while the default setup was used 
to specify the closed, core orbitals as $(6, 2, 2,0)$.
We used the the internally contracted MRCI algorithm (icMRCI)
implemented in MOLPRO.
The basis set used in our calculation
is aug-cc-pV$n$Z $n=3, 4, 5$ \cite{89Dunning.ai, 05BaPexx.ai}
so that we can estimate
the potential energy curve at the
complete basis set (CBS) limit by extrapolation.

According to Miliordos \etal,\cite{07MiMaxx.VO}
ionic avoided crossings are expected around 2.75 \AA,
while we found a discontinuity in the
dipole moment around 1.9 \AA.
We tried to add a second $^4\Sigma^-$ state
but failed to find an avoided crossing structure in that region.

Off-diagonal spin-orbit interaction between the
$\VOXSigma$ and $\VOAPi$ states
contributes to
the spin splitting of $\VOXSigma$.
As $\VOApPhi$ and $\VOAPi$ have the same
irreducible representations in the $C_{2v}$ group,
it is impossible to omit the $\VOApPhi$ in MRCI calculations.
Therefore,
we calculated the $\VOAPi$ and $\VOApPhi$ states together with
the $\VOXSigma$ states
using the same active space 
but only with the aug-cc-pVQZ basis set.

\subsection{Interaction of doublet states with $\VOXSigma$}

Previous studies \cite{jt623,jt644}
show that
the spin splitting of the $\VOXSigma$ state of VO is dominated by
the off-diagonal spin-orbit interaction
between its $\VOXSigma$ and $1\,^2\Sigma^+$ states.

The $1\,^2\Sigma^+$  state of VO,
designated $\mathrm{a}\,^2\Sigma^+$ in 
the experimental work of Adam \etal,\cite{95AdBaBeBo}
is easily obtained in a CASSCF calculation with MOLPRO
when its \texttt{LQUANT} 
(\ie the projection of orbit angular momentum on the internuclear axis)
is assigned.
However,
a MOLPRO MRCI calculation 
 may converge to the 
$1\,^2\Gamma$ state,
which has degenerate A$_1$ and A$_2$
representations.
The $1\,^2\Delta$ state also has the same irreducible representations
and is lower than
$1\,^2\Sigma^+$.
In principle,
the three states
$1\,^2\Sigma^+$,
$1\,^2\Gamma$
and $1\,^2\Delta$
should be optimized 
simultaneously in the
$^2$A$_1$ symmetry block.
Our calculation therefore included these three low-lying doublets  
states of VO
together with its ground state.
The two higher $^2\Pi$ states
were also included in the work of McKemmish \etal \cite{jt623}
but are not considered here.

We must provide a 
reasonable CASSCF reference 
for the MRCI calculations.
The $1\,^2\Sigma^+$ and
$1\,^2\Gamma$ states have the same
electron configuration as
X$\,^4\Sigma^-$
while $1\,^2\Delta$ has a
different one.\cite{09HoHaMa.VO}
Thus,
we initially calculated  only the $1\,^2\Delta$ and ground state, and
then subsequently added one   $^2\Gamma$ state and one $^2\Sigma^+$ state.
Nonetheless, 
we could not obtain the
correct $1\,^2\Delta$ state
in a state-average CASSCF calculation
including  $^4\Sigma^-$,
$^2\Gamma$,  $^2\Delta$
and $^2\Sigma^+$ 
when the closed orbitals were
set to $(6, 2, 2, 0)$.
To make the reference wavefunctions
physically appropriate,
we closed more orbitals,
 $(8, 2, 2, 0)$, in
CASSCF calculation,
while we still used the closed 
$(6, 2, 2, 0)$ space
in the subsequent icMRCI calculation.
Again we used an aug-cc-pVQZ basis set.

\subsection{Electron spin dipolar coupling 
and nuclear hyperfine coupling curves}

The electron spin-spin coupling  
was treated as an empirical fine tuning factor 
by  McKemmish \etal.\cite{jt644}
Using the quantum chemistry program ORCA,\cite{12ORCA.programs}
we calculated the electron spin-spin
dipolar contribution to the
zero-field splitting $\bm{D}$ tensor
of the ground state at the CASSCF level with
eleven electrons distributed in ten active orbitals.

Fully-resolved hyperfine splittings have been 
observed in the $v=0$ vibrational levels of the
$\VOXSigma$ state.
We calculated
the nuclear hyperfine $\bm{A}$ tensor
and the nuclear electric quadrupole coupling constant in ORCA,\cite{12ORCA.programs}
with the aim of predicting the  hyperfine 
structure in vibrationally-excited levels of VO.

The zero field splitting tensor
was calculated with an aug-cc-pVTZ basis set.
The nuclear magnetic $\bm{A}$-tensor and  
electric quadrupole coupling constant
were calculated with an aug-cc-pwCVQZ basis set.

The nuclear spin-rotation coupling constants
were calculated with another quantum chemistry program, DALTON \cite{14DALTON.method} 2020.0,
at the CASSCF level with an
aug-cc-pVQZ basis set.
The active space is the same as used in ORCA.

We failed to find a quantum chemistry program 
which calculates the electron spin-rotation constant $\gamma$ and therefore used 
the constant empirical value determined for $v=0$ instead
(See Table\,\ref{tab:vo_zfs_curve}).

\section{Ab initio results}

\subsection{$\VOXSigma$ potential energy curve}

\begin{figure*}
    \centering
    \includegraphics{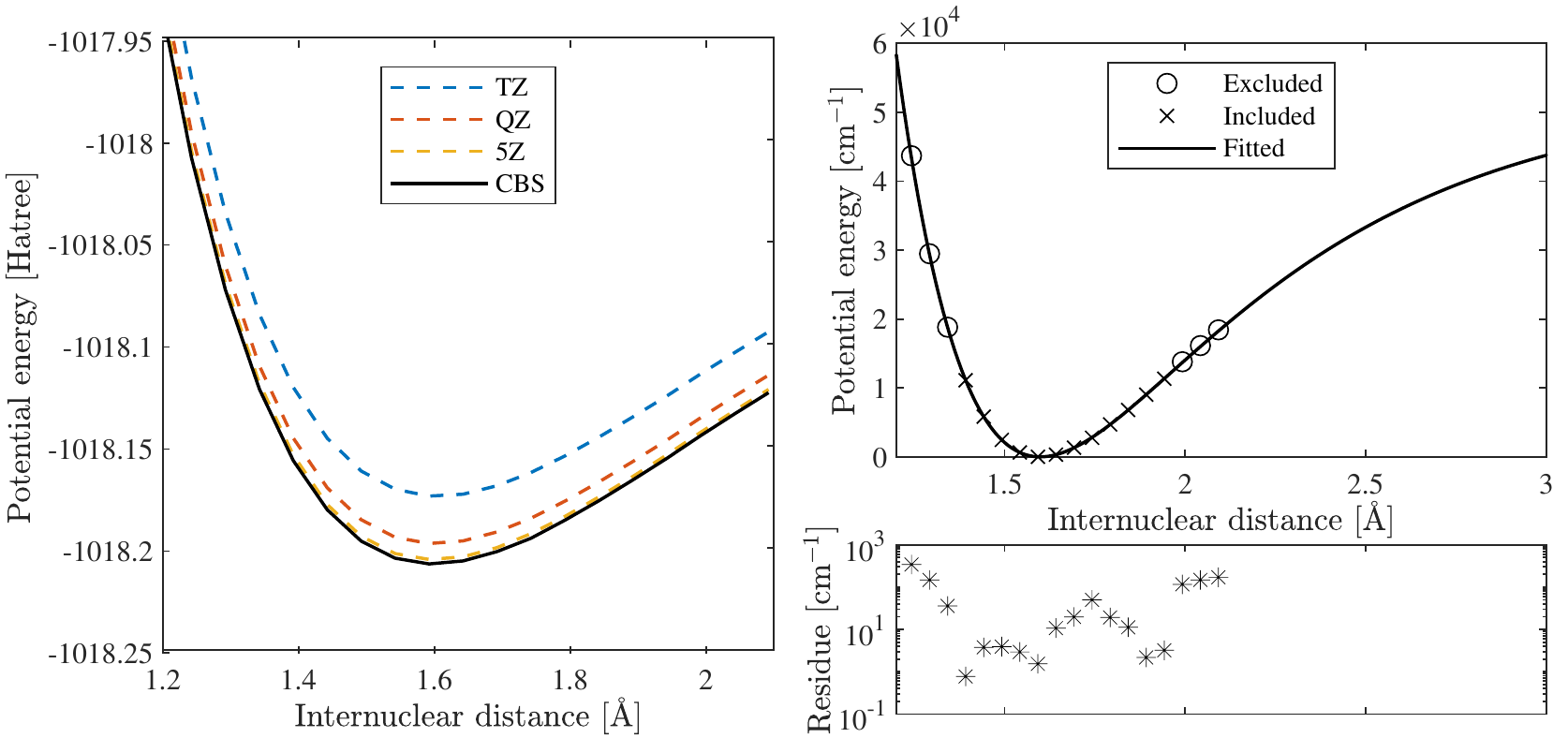}
    \caption{
    The lefthand panel shows 
    the MRCI+Q potential energy curves
    of the $\VOXSigma$ state
    calculated with aug-cc-pV$n$Z
    basis sets and
    the estimated one at
    complete basis set limit.
    The extrapolated potential energy
    curve is fitted with a second-order extended 
    Morse oscillator (EMO) function.
    The right-bottom panel shows the fitting residues.}
    \label{fig:vox_pec_extra}
\end{figure*}

The dashed curves in Fig.\,\ref{fig:vox_pec_extra}
are the \abinitio potentials
of the $\VOXSigma$ state of VO.
We estimated its potential energies
at the CBS limit using the formula
\[
E(n) = E_\mathrm{CBS} + \alpha\exp(-\beta n)
\]
and obtained the solid potential energy curve
shown in the left panel of 
Fig.\,\ref{fig:vox_pec_extra}.

The \abinitio curves were calculated 
to build the spectroscopic model of VO.
For numerical stability purposes,
we fitted the discrete points with continuous curves.
The extrapolated potential energy curve 
at the CBS limit
was fitted to a second-order extended Morse oscillator
(EMO) function:\cite{jt609}
\begin{equation}
    V(r)=T_{\mathrm{e}}+
    \left(A_{\mathrm{e}}-T_{\mathrm{e}}\right)
    \left[1-\exp \left(-\beta_{\mathrm{EMO}}(R)
        \left(R-R_{\mathrm{e}}\right)\right)
    \right]^{2},
    \label{eq:emo}
\end{equation}
where $R$ and $R_e$ is the internuclear distance
and its value at the equilibrium point and $A_{\mathrm{e}}$ is the asympotic energy relative to the minimum of the ground electronics state. 
$\beta_\mathrm{EMO}$ is expressed as
\begin{equation}
    \beta_{\mathrm{EMO}}(R)= b_0 + b_1\, y(R)
    +b_2\, y^{2}(R),
\end{equation}
where $y(R)$ is given by:
\begin{equation}
    y(R)=\frac{R^{4}-R_\mathrm{e}^{4}}
    {R^{4}+R_\mathrm{e}^{4}} \,.
    \label{eq:surkus}
\end{equation}
Only the points given as  crosses
in the righthand panel of Fig.\,\ref{fig:vox_pec_extra}
were included in the fit
to give a better approximation
of the lower vibrational levels.
Although the calculated potential energies
marked by circle 
were excluded,
they are still well represented by
the fitted curve.
The EMO parameters 
are listed in Table\,\ref{tab:fitted_pecs}.

\begin{figure}[h]
    \centering
    \includegraphics{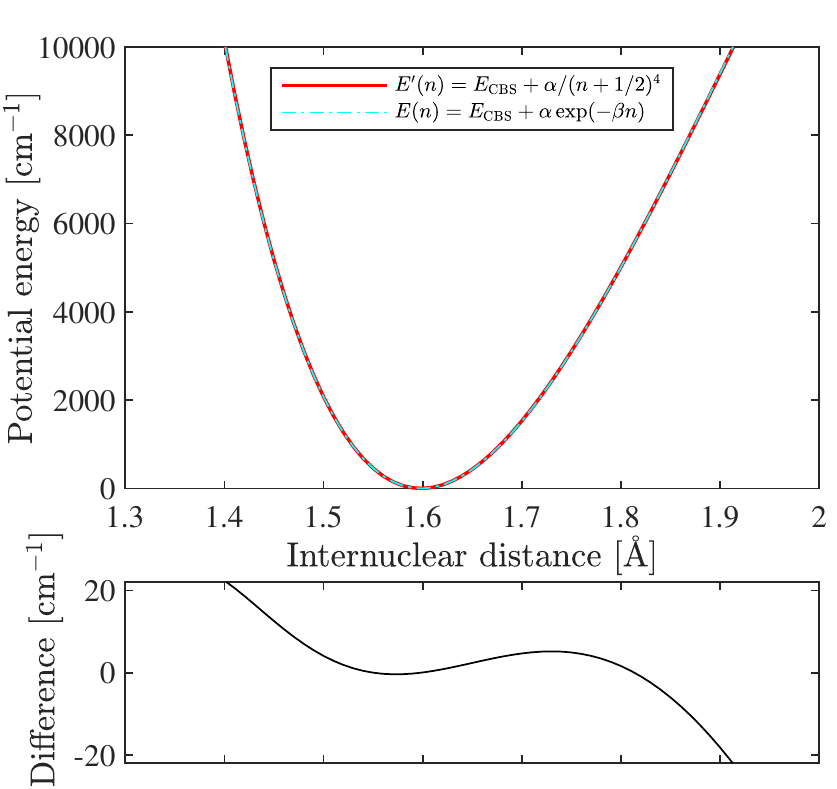}
    \caption{
    Fitted PECs corresponding to two 
    different extrapolation formulae
    as shown in the legend.
    The bottom panels show the 
    energy difference between the two curve.}
    \label{fig:compare_CBS}
\end{figure}

The fitted PEC is not sensitive to the extrapolation formula
in the region of interest (\ie $E\le \SI{10000}{\per\cm}$).
Figure\,\ref{fig:compare_CBS} compares 
the fitted EMO PECs of two extrapolation formulae:
$E'(n) = E_\mathrm{CBS} + {\alpha}/{(n+1/2)^4} $
and $E(n) = E_\mathrm{CBS} + \alpha\exp(-\beta n)$.
The EMO parameters corresponding to $E'(n)$ 
are listed in Table\,\ref{tab:fitted_pecs} too. 

\begin{table}
\centering
\caption{Optimized EMO parameters of the $\VOXSigma$ state.}
\label{tab:fitted_pecs}
\begin{ruledtabular}
\begin{tabular}{lrr}
Parameter & $E(n) = E_\mathrm{CBS} + \frac{\alpha}{(n+1/2)^4} $ & $E'(n) = E_\mathrm{CBS} + \alpha\mathrm{e}^{-\beta n}$ \\
\hline
$T_e$ [\si{\per\cm}]    & 0 & 0\\
$R_e$ [\si{\angstrom}]   & \num{1.598 438 63} & \num{1.5983 5533}    \\
$D_e$  [\si{\per\cm}]  & 52790 & 52790   \\
$b_0$ [\si{\per\angstrom}]  & \num{1.837 543 49}& \num{1.8404 2724}  \\
$b_1$ [\si{\per\angstrom}]  & \num{-9.6268 1017e-3}& \num{-1.6237 7024e-2}     \\
$b_2$  [\si{\per\angstrom}]  & \num{-1.4841 3484e-1} & \num{-1.8024 0476e-1}   \\
\end{tabular}
\end{ruledtabular}
\end{table}

\subsection{Potentials of $\VOAPi$ and $\VOdouSigmap$}

\begin{figure*}[h]
    \centering
    \includegraphics{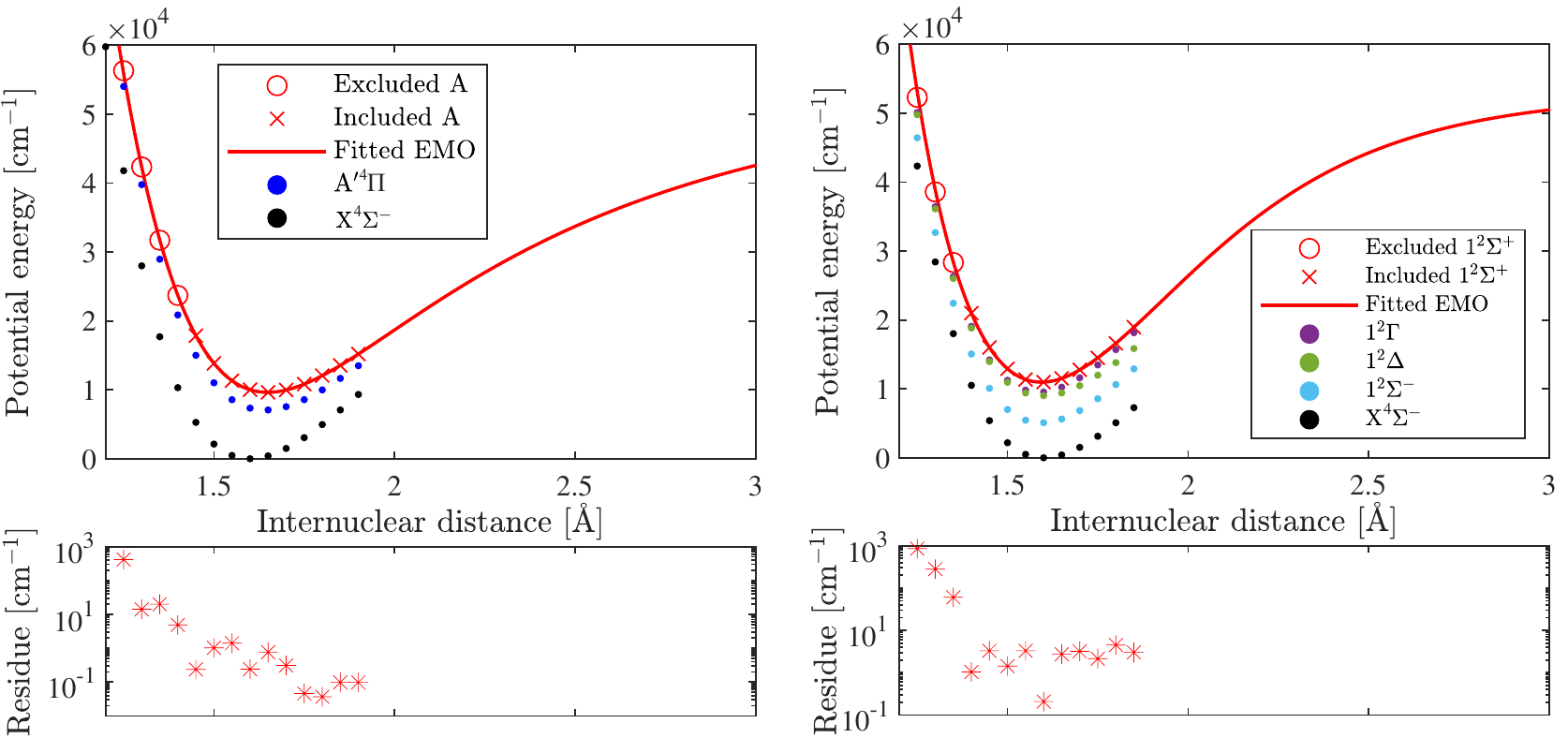}
    \caption{Calculated potential energy curves of the quartet states (left)
    and doublet states (right) of VO.
    The curves of $\VOAPi$ and $\VOdouSigmap$ were fitted with 
    EMO functions.
    The bottom panels show the fitting residues.}
    \label{fig:quartet_pec}
\end{figure*}

The calculated potential energy curves for the quartet
and doublet states are shown in Fig. \ref{fig:quartet_pec}.
The energies are shifted such that
the corresponding $\VOXSigma$ ground state of each set
has the same energy zero.
The potentials of $\VOAPi$ and $\VOdouSigmap$
were fitted with second-order EMO functions
whose parameters are listed 
in Table\,\ref{tab:fitted_excited_pecs}.

\begin{table}
    \centering
    \caption{Optimized EMO parameters of the excited states.}
    \label{tab:fitted_excited_pecs}
    \begin{ruledtabular}
    \begin{tabular}{lrr}
    Parameter & $\VOAPi$ & $\VOdouSigmap$\\
    \hline
    $T_e$ [\si{\per\cm}]    & \num{9.6344 5279e3} & \num{1.0973 9904e4}\\
    $R_e$ [\si{\angstrom}]    & \num{1.649 119 57}   & \num{1.593 447 21} \\
    $D_e$  [\si{\per\cm}]   & 52790   & 52790\\
    $b_0$ [\si{\per\angstrom}]  & \num{1.818 147 51}  & \num{2.126 803 13}\\
    $b_1$ [\si{\per\angstrom}]  & \num{-8.3538 5040e-2}   & \num{3.1622 7470e-1}  \\
    $b_2$  [\si{\per\angstrom}]   & \num{-3.1451 0129e-1}  & \num{1.9828 5641e-1} \\
    \end{tabular}
    \end{ruledtabular}
    \end{table}

\subsection{Spin-orbit couplings}

\begin{figure*}[h]
    \centering
    \includegraphics{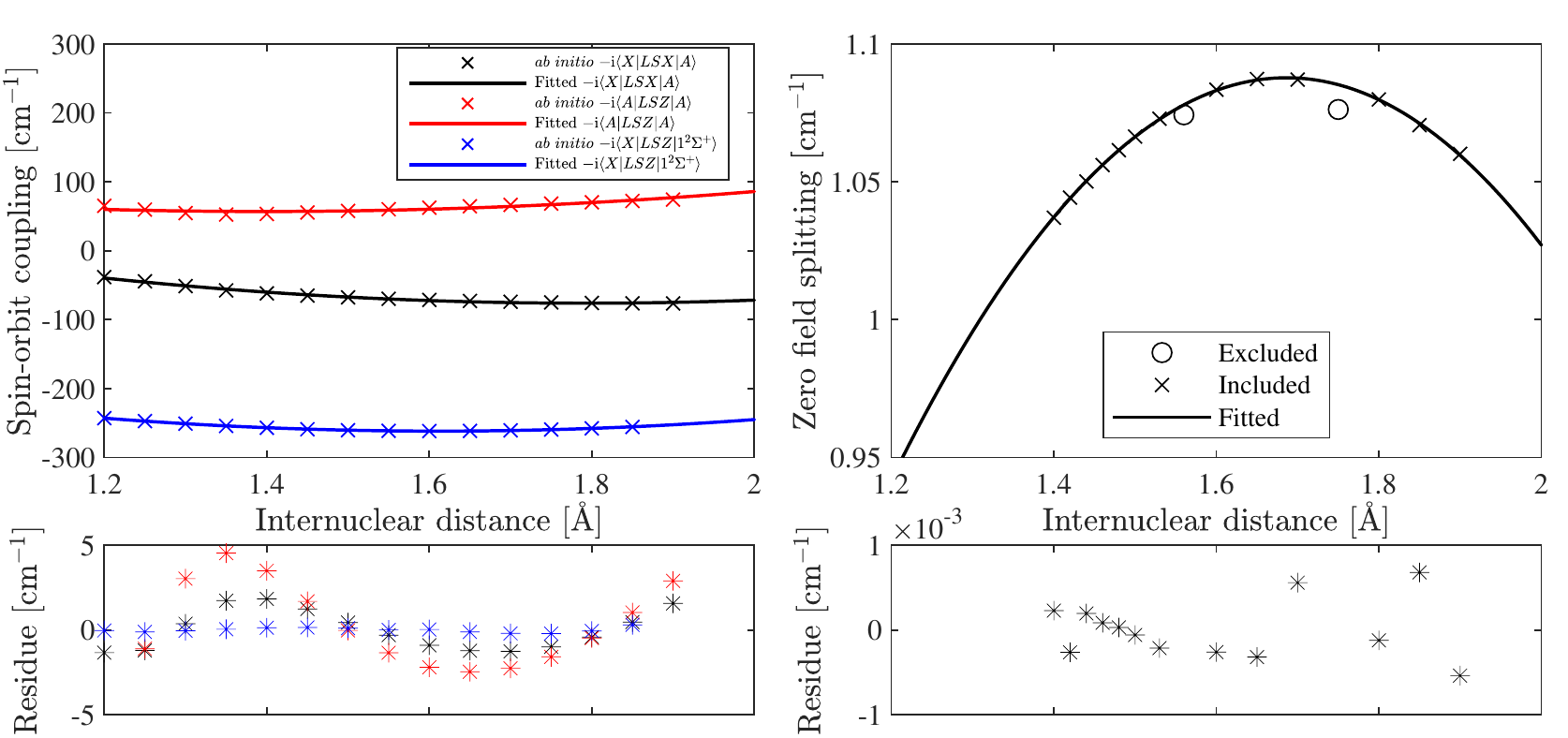}
    \caption{The calculated spin-orbit
    coupling curves (left) and 
    zero field splitting curve due to spin-spin coupling 
    (right) of VO
    which were fitted with polynomials.
    The bottom panels show the fitting residues.}
    \label{fig:VO_SOs}
\end{figure*}

The calculated spin-orbit coupling curves
are shown in the left panel of Fig.\,\ref{fig:VO_SOs}.
Note that the spin-orbit coupling constant
has a phase of $\iu$ as 
MOLPRO uses a Cartesian representation.
The figure demonstrates the real curves
multiplied an extra constant $-\iu$,
which were fitted with polynomials
\begin{equation}
    p(R) = \sum_i a_i\, R^i.
    \label{eq:poly}
\end{equation}
The polynomial coefficients $a_i$
are given in Table\,\ref{tab:vo_so_curves}.

\begin{table*}
\centering
\caption{Polynomial coefficients of 
the  \abinitio spin-orbit coupling curves. }
\label{tab:vo_so_curves}
\begin{ruledtabular}
\begin{tabular}{lrrr}
Coefficients & $-\iu\mel{\VOXSigma}{\mathcal{H}_\mathrm{LSX}}{\VOAPi}$ & $-\iu\mel{\VOAPi}{\mathcal{H}_\mathrm{LSZ}}{\VOAPi}$&$-\iu\mel{\VOXSigma}{\mathcal{H}_\mathrm{LSZ}}{\VOdouSigmap}$ \\
\hline
$a_0$ [\si{\per\cm}]   & \num{1.0420 0154e2} & \num{2.1166 1061e1}& \num{2.6259 8816e1}\\
$a_1$  [\si{\cm^{-1}\angstrom^{-1}}]   & \num{-3.7351 6108e2}& \num{-2.2176 9098e2}& \num{-3.5748 9537e2}\\
$a_2$  [\si{\cm^{-1}\angstrom^{-2}}]   & \num{2.5851 8247e2}& \num{7.9432 5083e1} & \num{1.1091 4862e2}\\
\end{tabular}
\end{ruledtabular}
\end{table*}

\subsection{Electron spin dipolar coupling}
In a Cartesian representation,
the zero-field splitting Hamiltonian is:\cite{01ScJexx.epr}
\begin{align}
    \opH_\mathrm{ZFS} &= \bmS^\transT\bm{D}\,\bmS.
\end{align}
where $\bmS = ( {S}_x, {S}_y, {S}_y)$ is the  spin vector operator and $\bm{D}$ is a dipolar interaction tensor. 
In  principle axes,
$\bm{D}$ is diagonal and 
\begin{equation}
    \opH_\mathrm{ZFS} = D_{xx}  {S}_x^2
    + D_{yy}  {S}_y^2 + D_{zz}  {S}_z^2.
\end{equation}
As a dipolar interaction tensor,
$\bm{D}$ is traceless
and thus $\opH_\mathrm{ZFS}$ only has two degrees of freedom.
In electron spin resonance spectroscopy,
it is usual to define two constants,
$D$ and $E$, to describe zero-field splitting:
\begin{align}
    D &= \frac{3}{2} D_{zz},\\
    E &= \frac{1}{2} (D_{xx} - D_{yy}).
\end{align}
The Hamiltonian can be rewritten as 
\begin{equation}
    \opH_\mathrm{ZFS} = 
    D\left[ {S}_z^2 -\frac{1}{3}\bmS^2\right]
    +E( {S}_x^2 -  {S}_y^2),
\end{equation}
with the principle axis chosen such that
\[
|E| \le \frac{1}{3} |D|.
\]
For the $\VOXSigma$ state of VO,  $E=0$,
and hence $D_{xx}=D_{yy}$.

The calculated zero-field splitting
curve is shown in 
the right panel of Fig.\,\ref{fig:VO_SOs}.
The two points marked by circles were excluded from the fit.
The other points were fitted with a parabolic curve
whose coefficients are given in
Table\,\ref{tab:vo_zfs_curve}.

We used the constant experimental value \cite{08FiZixx.VO}
for the spin-rotation coupling curve,
as shown in the last column of Table\,\ref{tab:vo_zfs_curve}.

\begin{table}
\centering
\caption{Polynomial coefficients of 
the \abinitio zero-field splitting curve
$D(R)$ and the empirical
spin-rotation curve $\gamma(R)$}
\label{tab:vo_zfs_curve}
\begin{ruledtabular}
\begin{tabular}{lrr}
Coefficients & $D(R)$ & $\gamma(R)$\\
\hline
$a_0$ [\si{\cm^{-1}}]   & \num{-6.6632 4020e-1} & \num{2.2421 1111e-2}\\
$a_1$ [\si{\cm^{-1}\angstrom^{-1}}]   & \num{2.0803 7245} \\
$a_2$  [\si{\cm^{-1}\angstrom^{-2}}]  & \num{-6.1684 6661e-1} \\
\end{tabular}
\end{ruledtabular}
\end{table}

\subsection{Nuclear hyperfine couplings} 

In a Cartesian representation,
the Hamiltonian for  nuclear spin --
electron spin magnetic interaction
is:\cite{01ScJexx.epr}
\begin{align}
    \opH_\mathrm{HFS} &= \bmS^\transT\bm{A}\,\bmI.
\end{align}

\begin{figure*}
    \centering
    \includegraphics{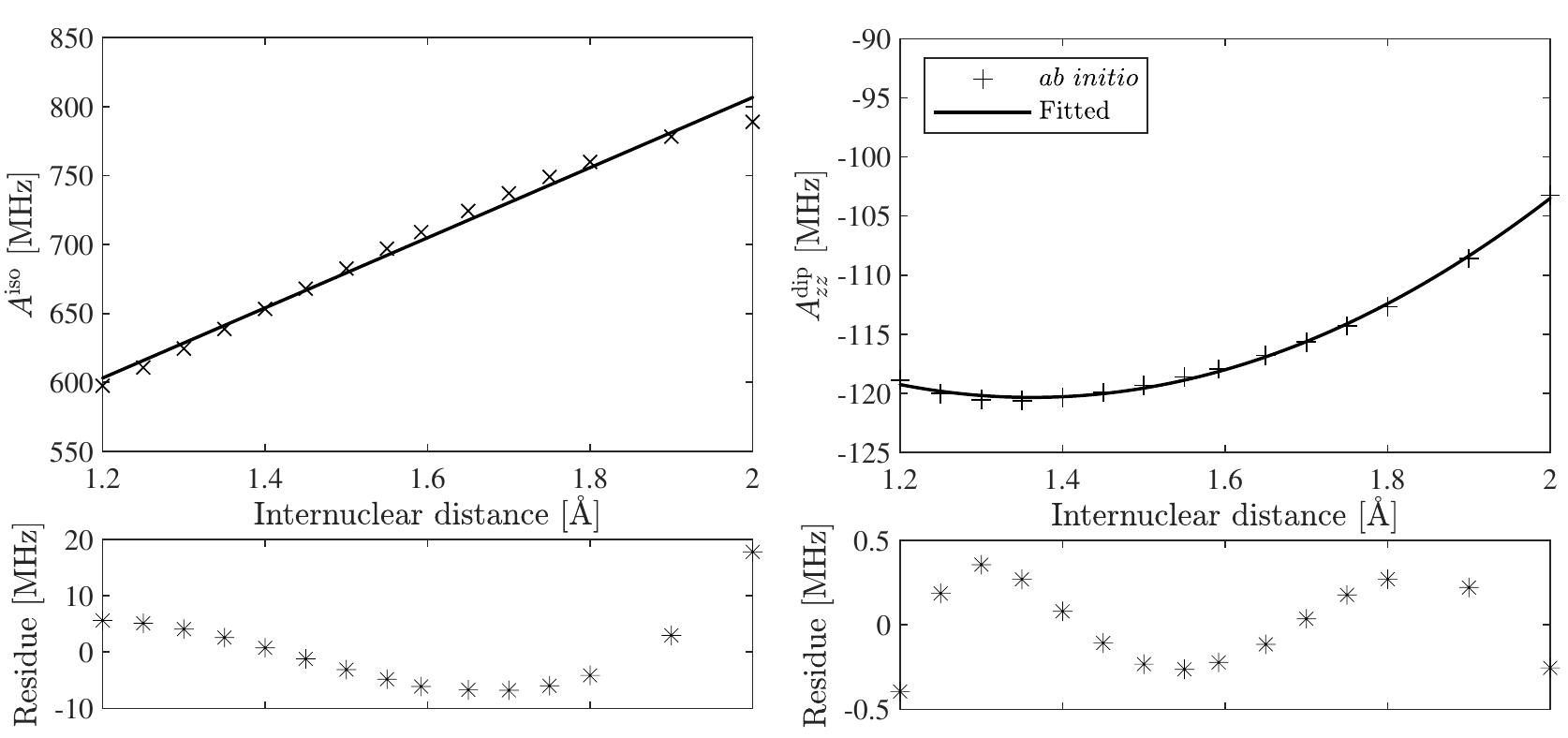}
    \caption{The calculated $A^\mathrm{iso}$ 
    and $A^\mathrm{dip}_{zz}$ curves of $\VOXSigma$
    which were fitted with polynomials.
    The bottom panels show the fitting residues.}
    \label{fig:vox_fc}
\end{figure*}

The hyperfine coupling tensor
can be divided into an isotropic term $A^\mathrm{iso}$
and a dipolar term $\bm{A}^\mathrm{dip}$:
\begin{equation}
    \opH_\mathrm{HFC} 
    =A^\mathrm{iso}\,\bmS\cdot \bmI +
    \bmS^\transT \bm{A}^\mathrm{dip}\, \bmI.
\end{equation}
$A^\mathrm{iso}$ is also known as 
the Fermi-contact interaction constant.
The isotropic hyperfine coupling constant is given by
\begin{equation}
   A^\mathrm{iso} = \frac{1}{3}
   \left(A_{xx}+A_{yy} +A_{zz}
   \right).
\end{equation}
The calculated curve ${A}^\mathrm{iso}$
are shown in the left panel of Fig.\,\ref{fig:vox_fc}.
The points were fitted with a linear function,
whose coefficients are given in Table\,\ref{tab:vo_hfc_curves}.

In the principle axis representation,
the off-diagonal matrix elements of 
the dipolar interaction tensor $\bm{A}^\mathrm{dip}$ 
vanish.
Since 
$\bm{A}^\mathrm{dip}$
is also traceless, 
we obtain
\begin{equation}
   A_{xx}^\mathrm{dip}+A_{yy}^\mathrm{dip} +A_{zz}^\mathrm{dip}=0.
\end{equation}
Moreover,
\begin{equation}
   A_{xx}^\mathrm{dip}=A_{yy}^\mathrm{dip},
\end{equation}
for the $\VOXSigma$ state.
Thus, 
there is only one independent parameter for $\bm{A}^\mathrm{dip}$.
The calculated $A_{zz}^\mathrm{dip}$ 
term, which is  plotted in the right panel of Fig.\,\ref{fig:vox_fc},
was fitted with a parabolic curve
whose coefficients are given in Table\,\ref{tab:vo_hfc_curves}.

\begin{table*}
\centering
\caption{Polynomial coefficients of the \abinitio hyperfine coupling curves
of the $\VOXSigma$ state.}
\label{tab:vo_hfc_curves}
\begin{ruledtabular}
\begin{tabular}{lrrrr}
Coefficients & ${A}^\mathrm{dip}_{zz}$ & ${A}^\mathrm{iso}$&$eQq_0$ &$c_I$\\
\hline
$a_0$ [\si{MHz}]   & \num{-4.35374634e+1}  & \num{ 2.95222135e+2}&\num{ -3.67214582e+03}  &\num{3.77322818e+04}\\
$a_1$ [\si{MHz\,\angstrom^{-1}}]   &\num{-1.12799291e+2}   &\num{2.56635489e+2}  &\num{1.00349024e+04}  &\num{-1.31588701e+05}\\
$a_2$ [\si{MHz\,\angstrom^{-2}}]   &\num{4.14063843e+01}  &  &\num{-1.09166936e+04}  &\num{1.82538509e+05}\\
$a_3$ [\si{MHz\,\angstrom^{-3}}]  &   & &\num{5.91507164e+03} &\num{-1.26234203e+05}\\
$a_3$ [\si{MHz\,\angstrom^{-4}}]  &   & &\num{-1.60068427e+03} &\num{4.35197409e+04}\\
$a_4$ [\si{MHz\,\angstrom^{-5}}]  &   & &\num{1.73355438e+02 }&\num{-5.99532161e+03}\\
\end{tabular}
\end{ruledtabular}
\end{table*}

\begin{figure*}
    \centering
    \includegraphics{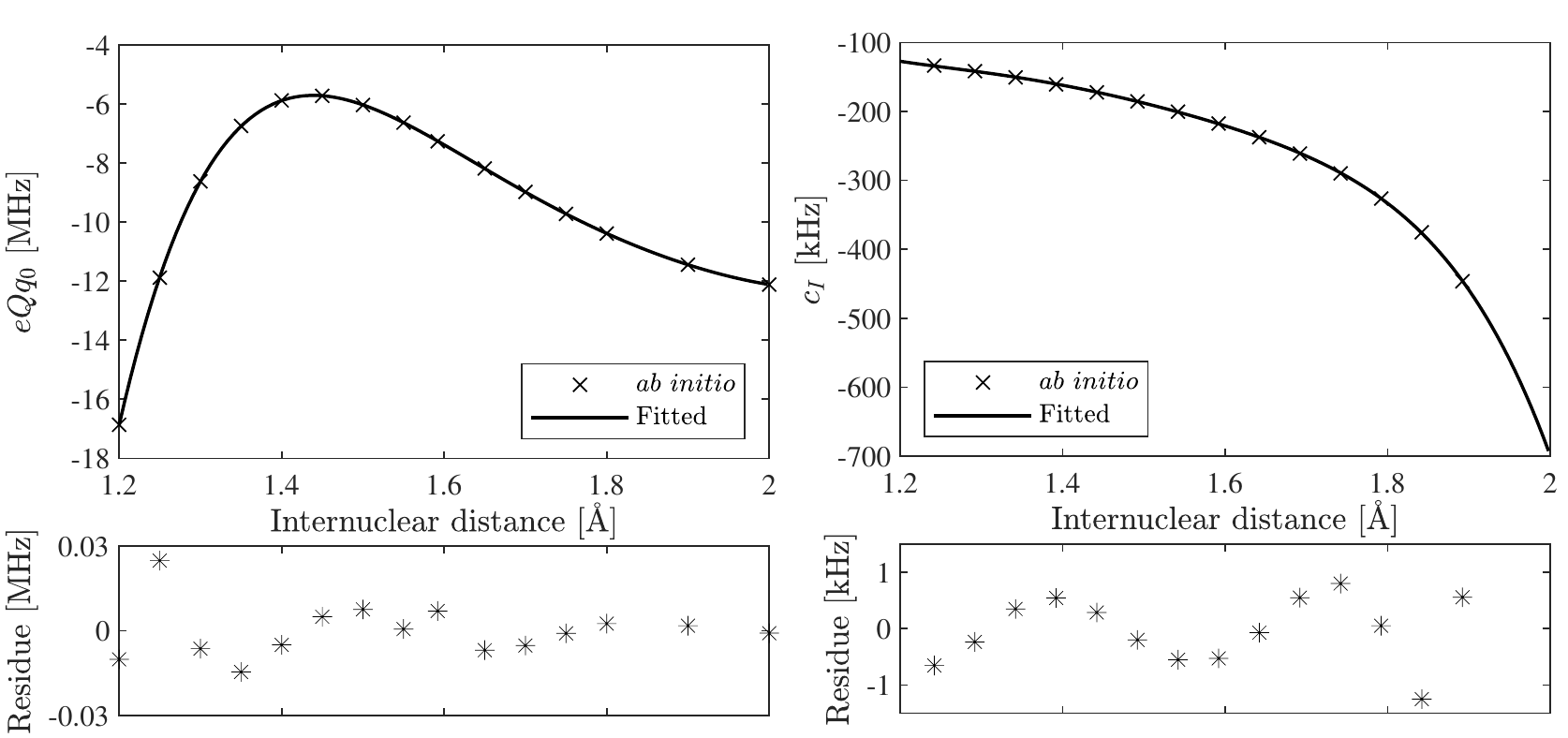}
    \caption{The calculated 
    nuclear electric quadrupole and
    nuclear spin-rotation coupling
    curve of $\VOXSigma$
    which were fitted with polynomials.
    The panels show the fitting residues.}
    \label{fig:vox_eqq}
\end{figure*}

The nuclear electric quadrupole 
coupling and
nuclear spin-rotation coupling
are relatively weak for the $\VOXSigma$ state
as shown in Fig.\,\ref{fig:vox_eqq}.
They were fitted by polynomials, see eq.~\ref{eq:poly},
whose coefficients are listed in
Table\,\ref{tab:vo_hfc_curves}.

\section{Infrared spectra}

\subsection{Spectroscopic model}
A spectroscopic model 
considering the $\VOXSigma$,
$\VOAPi$ and $\VOdouSigmap$ states
of VO was developed for
the diatomic variational nuclear motion program \duo.\cite{jt609}
The 
equilibrium bond length of
the $\VOXSigma$ \abinitio PEC
was shifted by about \SI{0.009}{\angstrom}
so that
\begin{equation}
    R_e = 1.5894809\ \si{\angstrom},
\end{equation}
resulting in the correct rotational constant.

For the basis set in \duo we used 20 vibrationally contracted basis functions
for the ground electronic states and 10 for the other two electronic states
based on 401 sinc-DVR grid points,
covering the internuclear distance range
from 1.2 to 4 \si{\angstrom}. 
The upper limit of the energy calculations
was set to \SI{50000}{\per\cm},
which is just below
the first dissociation limit of VO;
the  energy levels of interest for this work are expected to be below \SI{10000}{\per\cm} 
which is close to the 
$T_e$ value of $\VOAPi$
and is also below the discontinuity point in the PEC of the $\VOXSigma$ state.
This range covers  vibrational levels up to $v= 10$.
Thus, the 20 vibrational contracted basis functions
are enough to give converged energy levels.

The coupling constants used in \duo
follow the definitions generally adopted 
in experimental studies.\cite{jt855}
Some constants have the same definition as
those given by quantum chemistry programs.
For example,
the Fermi-contact coupling constant 
is just ${A}^\mathrm{iso}$ 
\begin{equation}
    b_\mathrm{F} = A^\mathrm{iso}.
\end{equation}
Definitions of others are different
and we give the relevant interconversion 
formulae below.

In a Cartesian representation,
the Hamiltonian of a diagonal electron spin-spin dipolar interaction of diatomic molecule is 
\begin{equation}
    \opH_\mathrm{SS} = 
    \frac{2}{3} \lambda \left(3  {S}_z^2 - \bmS ^2\right),
\end{equation}
where $\bmS$ is the electron spin angular momentum
and $S_z$ is its $z$ component.
Comparing $\opH_\mathrm{SS}$ with $\opH_\mathrm{ZFS}$,
we have
\begin{equation}
    \lambda = \frac{1}{2}D.
\end{equation}

In a Cartesian representation,
the Hamiltonian of the
nuclear spin-electron spin
dipolar  interaction is given by
\begin{align}
    \opH_\mathrm{dip} &=
    \frac{1}{3} c\left(3  {I}_{z}  {S}_{z}-\bmI \cdot \bmS\right) \notag\\
    &+ \frac{1}{2} 
    d\left[ {S}_{+}  {I}_{+} 
    \exp(-2 \iu \phi)+ {S}_{-}  {I}_{-} 
    \exp(2 \iu \phi)\right] \notag\\
    &
    e\Bigg[\left( {S}_{-}  {I}_{z}+ {S}_{z}  {I}_{-}\right) \exp(\iu \phi)\notag\\
    &\quad+\left( {S}_{+}  {I}_{z}+ {S}_{z}  {I}_{+}\right) \exp(-\iu \phi)\Bigg],
\end{align}
where $c$, $d$ and $e$ are three nuclear spin-electron spin
dipolar interaction constants;
$\bmI$ is the nuclear spin angular momentum;
$I_z$, $I_+$ and $I_-$ are the components of $\bmI$;
$S_z$, $S_+$ and $S_-$ are the components of $\bmS$;
$\phi$ is the variable of spherical harmonics,
see Eq.\,4 of \citet{95SlClJa.hyperfine}.
Comparing the Hamiltonian with the matrix elements
of $\bmI^\transT \bm{A}^\mathrm{dip}\, \bmS$,
we have:
\begin{align}
    A_{xx}^\mathrm{dip} &= -\frac{c}{3}+d \cos(2\phi),\\
    A_{yy}^\mathrm{dip}  &= -\frac{c}{3}-d \cos(2\phi),\\
    A_{zz}^\mathrm{dip}  &= \frac{2c}{3}.
\end{align}
For the ground state,
we have $A_{xx}^\mathrm{dip} =A_{yy}^\mathrm{dip}$.
The only non-vanishing constant is 
\begin{equation}
    c=\frac{3}{2}A_{zz}.
\end{equation}

Dipole moments were also obtained 
from our \abinitio\ calculations.
However,
they are not as accurate
as the dipole moments calculated
by McKemmish \etal \cite{jt623}
using the finite-field method.
Thus, we used the permanent dipole moment of $\VOXSigma$ in 
Ref.\cite{jt623} to compute 
Einstein-$A$ coefficients and hence transition intensities.

\subsection{Hyperfine matrix elements 
in the representation of the vibrational basis set}

\begin{figure}[h]
    \centering
    \includegraphics{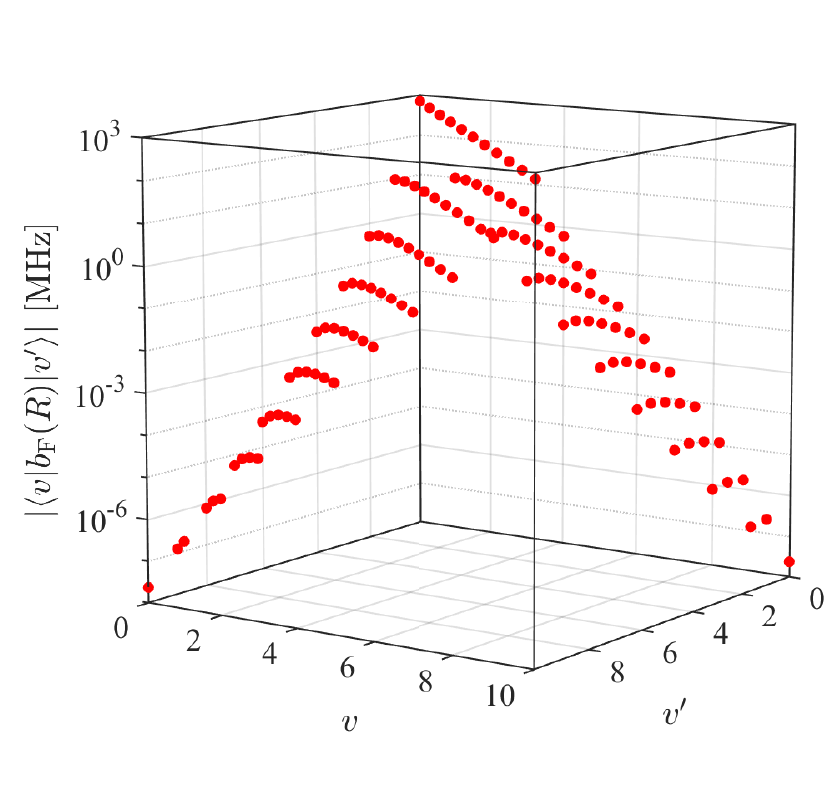}
    \caption{Absolute values of Fermi-contact
    matrix elements 
    $\mel{v}{b_\mathrm{F}(R)}{v'}$
    of $\VOXSigma$
    for $v\leq 10$
    and $v'\leq 10$.
    }
    \label{fig:fc_mel}
\end{figure}

We use a fully variational method to 
calculated the hyperfine structure of the VO $\VOXSigma$ state.
The final wavefunctions
have non-zero projections on
all contracted vibrational basis functions.
See our previous paper \cite{jt855} for more details.
The absolute values of the Fermi-contact
matrix elements 
$\mel{v}{b_\mathrm{F}(R)}{v'}$
are plotted in Fig.\,\ref{fig:fc_mel}.
The values decrease dramatically with
the difference between $v'$ and $v$,
\ie, the diagonal matrix element
$\mel{v}{b_\mathrm{F}(R)}{v}$
dominates the Fermi-contact interaction in the vibrational states.
The reason for the phenomenon is that
the lowest 11 vibrational levels of $\VOXSigma$
do not interact with other vibronic levels in our model.
Thus, the diagonal Fermi-contact matrix  elements in the VO $\VOXSigma$ state provided
should be equivalent to the  spectroscopic 
coupling constants used in effective Hamiltonian methods. We list all the diagonal hyperfine matrix elements of the lowest
11 vibrational levels
of $\VOXSigma$ in Table\,\ref{tab:hyper_constants}.
Compared to the measured
constants of the $v=0$ level\cite{08FiZixx.VO},
the absolute values of the calculated Fermi-contact matrix elements
are smaller 
while the nuclear spin-electron spin 
dipolar matrix elements are larger.
For VO, the nuclear spin-rotation and
nuclear electric quadrupole interactions
are much weaker than the other hyperfine interactions.
The corresponding matrix elements
are of similar magnitude to the experimental values.

\begin{table}    
\caption{The empirical hyperfine coupling constants
for $v=0$
given in Table\,4 of \citet{08FiZixx.VO} and 
the calculated diagonal hyperfine
matrix elements $\mel{v=0}{\cdot}{v=0}$ 
of $\VOXSigma$.
All values are given in MHz.
}
\label{tab:hyper_constants}
\begin{ruledtabular}
\begin{tabular}{ccccc}
Parameter  & ${b_\mathrm{F}}$ & ${c}$ & ${c_I}$ & ${eQq_0}$\\
\hline
Empirical\cite{08FiZixx.VO} &778.737(66) & -129.84(19) &0.1928(51) &-2.5(1.3)\\

 $Ab\ initio$    & 703.2540 & -177.1301 & -0.2191 & -7.2987 \\
\end{tabular}
\end{ruledtabular}
\end{table}

\subsection{Hyperfine eigenstates and transitions}

A hyperfine-resolved line list was  generated
based on the spectroscopic model. 
\duo provides data in ExoMol format\cite{jt548} which means 
energies with quantum numbers
are in a \texttt{.states}  file
and the Einstein-A coefficients for each
transitions are in a \texttt{.trans} file.
Examples of calculated energies and transitions
extracted from the output files are
given as supplementary materials.
In \duo's outputs,
the eigenstates are printed in the
increasing order of the final angular momentum,
which is $F$ here.
All energies are given relative to
the non-hyperfine zero-point energy
\ie, which corresponds to 
$J=0.5$, $+$ parity, and $v=0$.

\begin{figure*}
    \centering
    \includegraphics{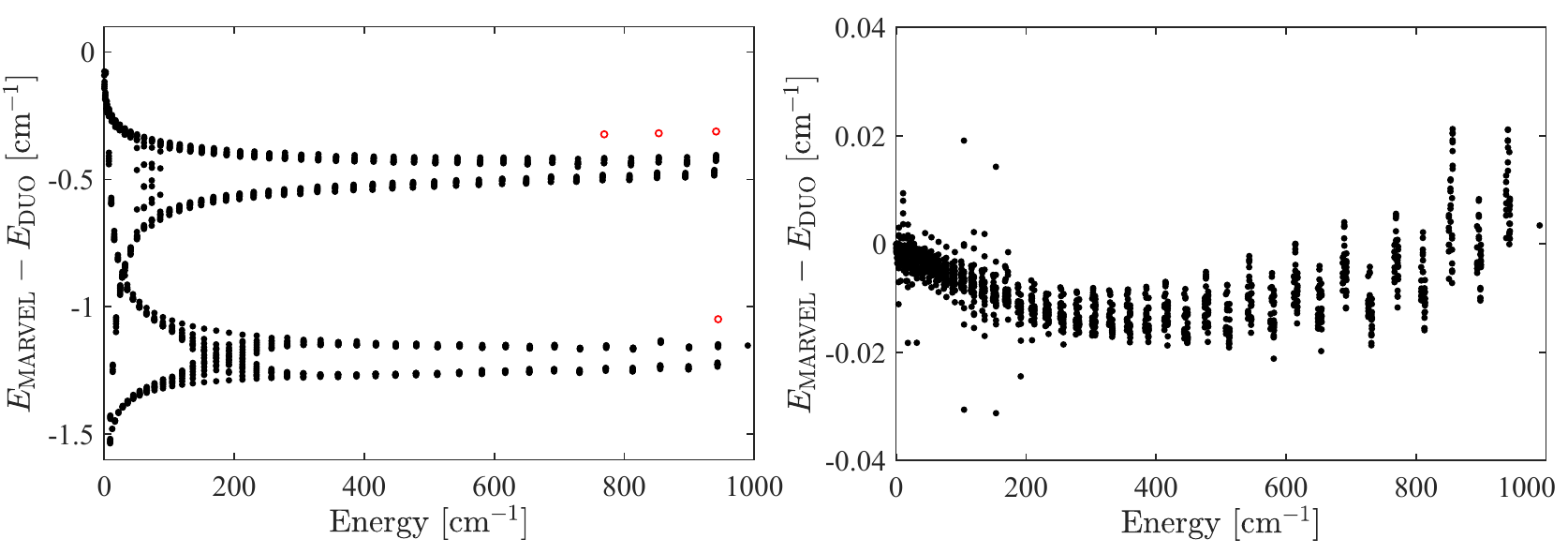}
    \caption{Energy differences between 
    results of \duo and MARVEL analysis
    when using \abinitio curves.
    Left: only the $R_e$ value 
    of $\VOXSigma$ was shifted
    to give correct rotational constants.
    Right:
    several other curves were also shifted to 
    reproduce the coupling constants
    given in Table\,4 of \citet{08FiZixx.VO}.
    }
    \label{fig:fig_VOX_duo_energy}
\end{figure*}

A hyperfine-resolved set of empirical energies of  VO has 
recently  been obtained \cite{jt869} using the MARVEL 
(measured active vibration-rotation
energy levels) procedure,
which
includes 6603 validated transitions
from three experimental sources
\cite{82ChHaMe,95AdBaBeBo, 08FiZixx.VO}
and 
gives
1256 hyperfine-resolved 
energy terms for the $v=0$
state of $\VOXSigma$.
We compare our calculated energies
with all  the MARVEL ones, as illustrated in
the left panel of Fig.\,\ref{fig:fig_VOX_duo_energy}.
The energy differences indicate that the \abinitio fine 
and hyperfine coupling curves
require further refinement to 
give accurate electron and nuclear spin splitting.

In order to illustrate the potential of such
refinement on the quality of the energy calculations, 
we shifted some fine and 
hyperfine coupling curves in our model
such that the corresponding diagonal matrix elements $\mel{v=0}{\cdot}{v=0}$
have the same values as the 
experimental spectroscopic constants 
determined by \citet{08FiZixx.VO}. 
The shifted parameters are listed in Table\,\ref{tab:shifted_a0}.
The right panel of 
Fig.\,\ref{fig:fig_VOX_duo_energy}
demonstrates the differences between the 
calculated and MARVEL energies in this case.
The calculation accuracy improves significantly with use of the shifted curves.

\begin{table}[h]
    \caption{
    Final $a_0$ values for four shifted 
    curves:  spin-orbit interaction of $\VOXSigma$-$\VOdouSigmap$;
    $\gamma(R)$,
    $A^\mathrm{iso}(R)$ and $A_{zz}^\mathrm{dip}(R)$ of $\VOXSigma$.
    }
    \label{tab:shifted_a0}
    \begin{ruledtabular}
    \begin{tabular}{llr}
    Curve & State(s)&$a_0$ value  [\si{\per\cm}]\\
    \hline
    Spin-orbit &$\VOXSigma -\VOdouSigmap$ & \num{5.9683 4165e1}\\
    $\gamma(R)$  &$\VOXSigma$ & \num{2.2181 1385e-2}\\
    $A^\mathrm{iso}(R)$  &$\VOXSigma$& \num{1.2709 6358e-2}\\
    $A_{zz}^\mathrm{dip}(R)$ &$\VOXSigma$  & \num{-4.0066 1787E-4} \\
    \end{tabular}
    \end{ruledtabular}
\end{table}

\begin{figure*}
    \centering
    \includegraphics{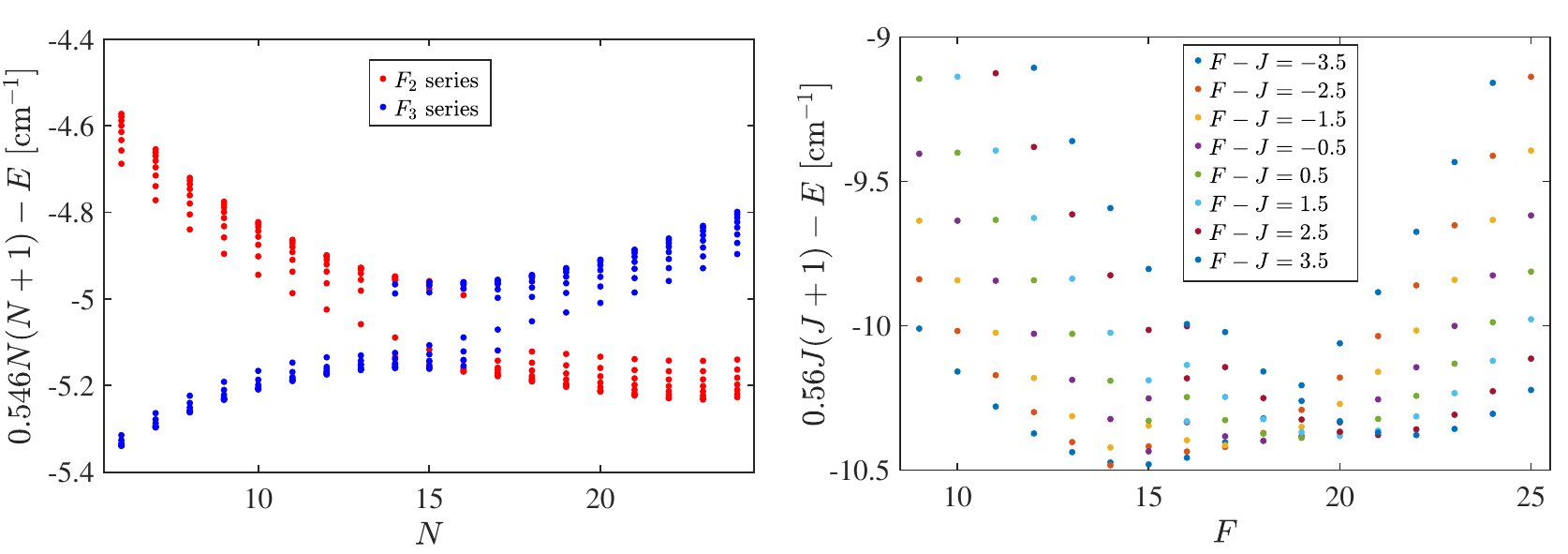}
    \caption{
    The lefthand panel shows 
    the avoided crossing structure of the 
    $F_2$ and $F_3$ levels of $\VOXSigma$.
    The righthand panel shows the mixing energy levels in 
    the $F_2$ series of $\VOXSigma$.    
    Note that,
    \duo\ does not use the quantum number $N$. $N$ is given 
    here simply for the clarity of the figure and was obtained
    using the  rule
    $N=J-0.5$ for the $F_2$ series and
    $N=J+0.5$  for the $F_3$ series.
    }
    \label{fig:vox_hyperfine_interaction}
\end{figure*}

There are four states
(shown as red circles in the lefthand panel)
whose calculation errors are greater than \SI{0.1}{\per\cm}, so
outside the range of the righthand panel of Fig.\,\ref{fig:fig_VOX_duo_energy}.
The energy levels between 100 to 200 \si{\per\cm}
have larger uncertainties than the others,
as shown in the right panel.
As discussed previously,\cite{82ChHaMe,95AdBaBeBo, 08FiZixx.VO}
this behavior arises from the internal  perturbations
near $N= 15$,
resulting in an avoided crossing structure 
as shown the lefthand panel of Fig.\,\ref{fig:vox_hyperfine_interaction}.
The righthand panel of Fig.\,\ref{fig:vox_hyperfine_interaction} illustrates the 
interactions of states in the $F_2$ series of $\VOXSigma$.
The interactions mix energy levels
which makes it difficult to 
assign quantum number to these states.
The globally $J$-dependent 
systematic error 
can be attributed to  inaccurate 
spin-orbit, spin-spin and spin-rotation
coupling curves.
We plan to refine these curves
in our future work.

\subsection{Transition intensities and lifetime}

\begin{figure*}[h]
    \centering
    \includegraphics{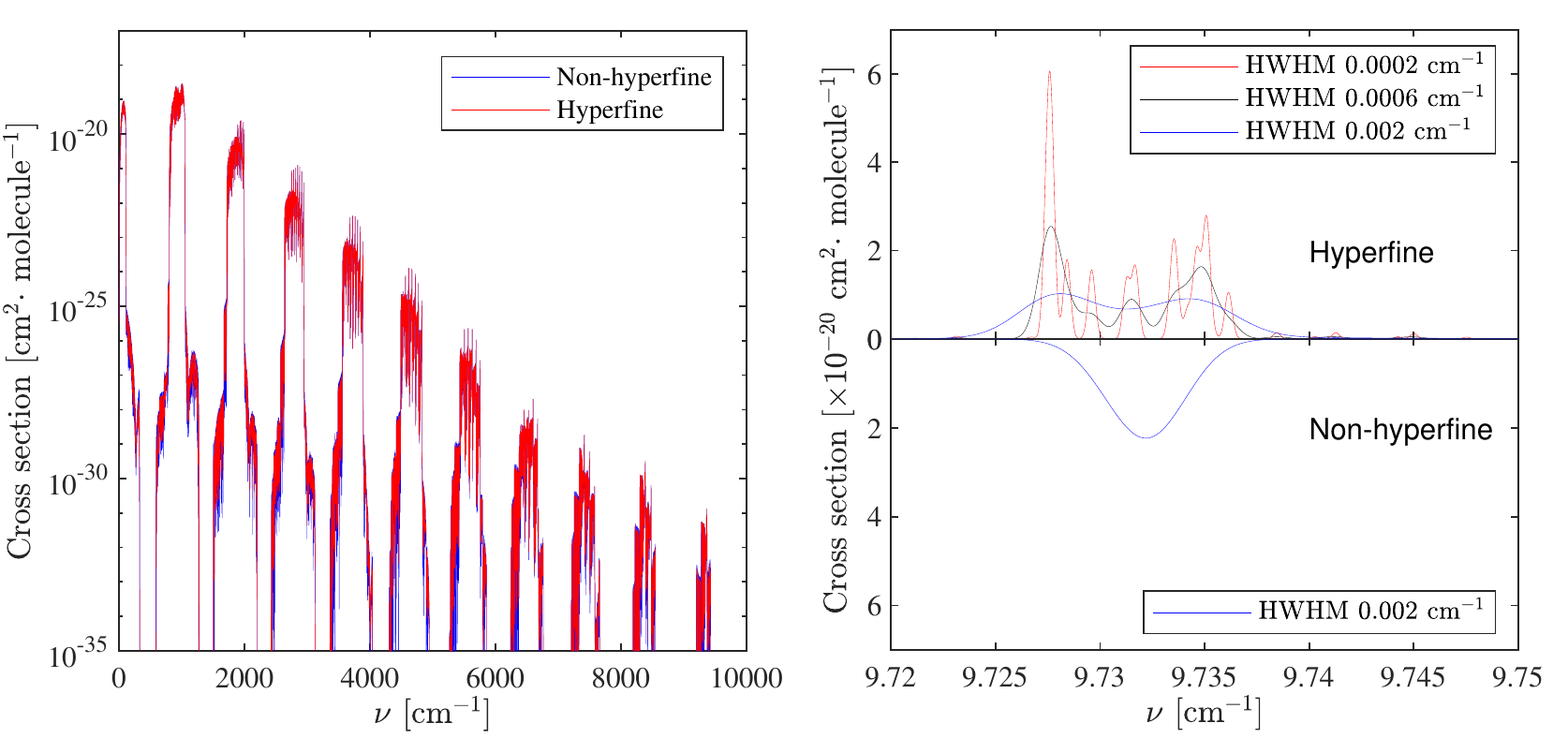}
    \caption{Comparison of 
    VO IR cross sections at 2200 K.
    Left:
    the cross sections were calculated
    with Gaussian profiles whose linewidth are \SI{0.2}{\per\cm}.
    Right:
    the cross sections were calculated with
    Gaussian profiles of different line widths in a narrow range.
    `Non-hyperfine' in this and following figures is a short notation
    which means that
    the spectra were simulated without
    considering nuclear hyperfine couplings.}
    \label{fig:compareVOcross}
\end{figure*}

The hyperfine resolved VO line list was used to generate spectra 
of the $\VOXSigma$ band using the  program   \textsc{ExoCross}.\cite{jt708}
The left panel of Fig.\,\ref{fig:compareVOcross}
compares cross sections calculated in this work
at $T= 2200$~K.
We used a Gaussian lineshape function for each 
isolated line and the linewidth was chosen as
\SI{0.2}{\per\cm}.
The linewidth is wider than hyperfine splittings 
and thus, the cross section profiles are blended.
As a result, the hyperfine resolved and unresolved cross sections
agree well with each other.
Note that, 
in this work, we only calculated the 
transitions within the ground state of VO without 
considering the A-X transition dipole moment contribution 
to line strengths.
In practice,
A-X spin-orbit coupling
mixes the wavefunctions of the two electronic states
meaning spectra are increasingly determined
by both the X-X and the A-X electric dipole moment curves;
transitions
above \SI{6000}{\per\cm}
are much stronger when
the A-X transition dipole moment is included.
We do not attempt
to properly model the A state here so we leave the discussion of 
the interaction of this and other electronic states to  future work.

Only hyperfine transitions with narrow broadening
parameters are distinguishable 
in high-resolution experiments, 
see \eg the work of \citet{08FiZixx.VO}.
We simulated the spectra of the eight hyperfine transitions near \SI{9.73}{\per\cm} 
with different line widths.
As shown in the top-right panel of Fig.\,\ref{fig:compareVOcross},
the hyperfine transitions are completely blended
when the half width at half maximum is \SI{0.002}{\per\cm}.
However,
due to the uneven line strength distribution of hyperfine transitions,
the shape and center of the blended profile differs from 
the one simulated from the line list
without considering the nuclear hyperfine couplings,
which is shown in the bottom-right panel of Fig.\,\ref{fig:compareVOcross}.
Similar conclusions were drawn from the VO MARVEL study \cite{jt869} where
attempts to deperturb the hyperfine-resolved energies by setting the hyperfine constants to zero were found to give poor results.

\begin{figure*}
    \centering
    \includegraphics{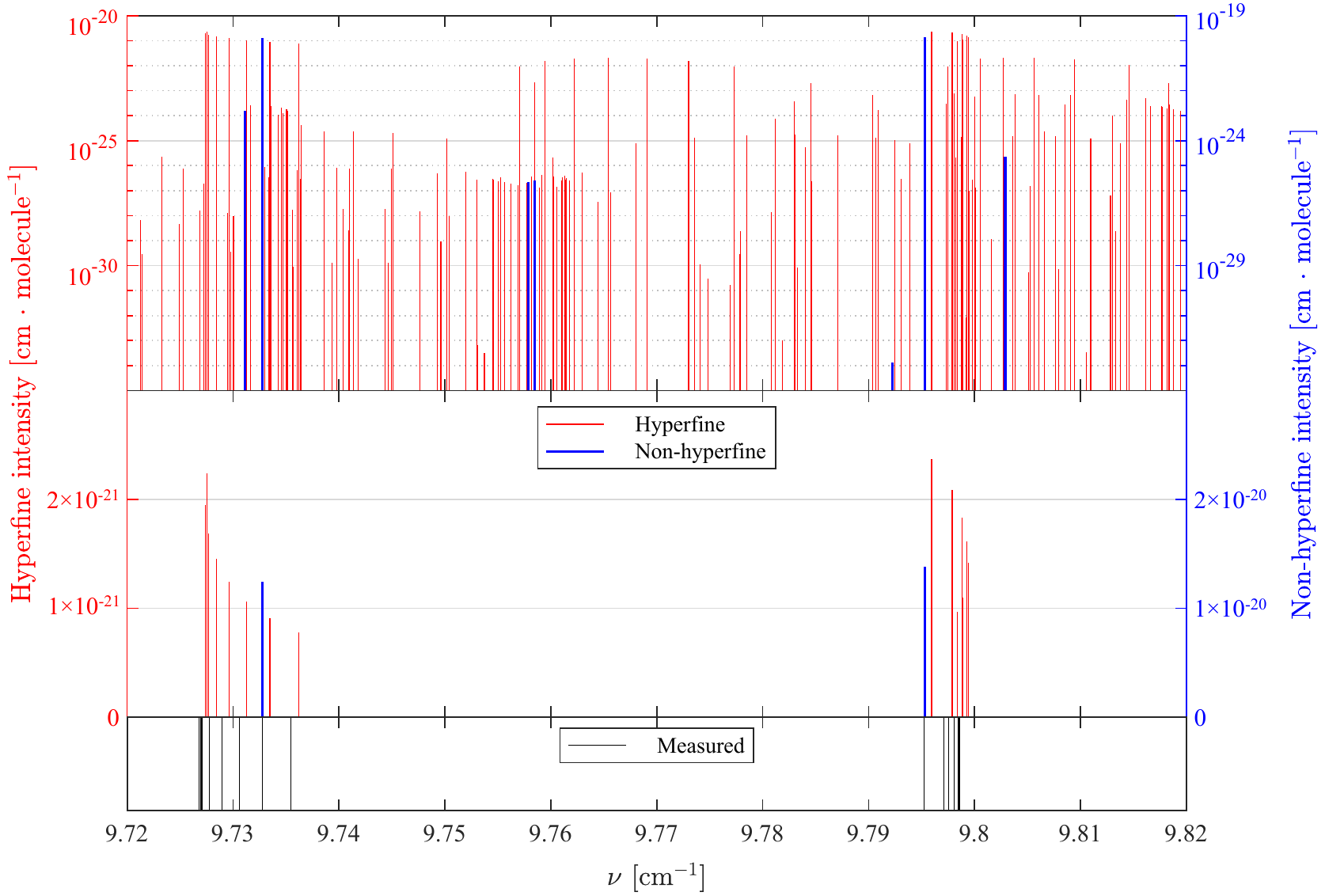}
    \caption{
    Comparison of the 
    calculated (top and middle)
    and measured (bottom) transitions 
    near \SI{9.77}{\per\cm}.
    The line intensities in the top
    and middle panels
    were calculated at \SI{208}{K}.
    The middle panel only demonstrates
    the strong transitions.
    The hyperfine resolved 
    line positions in the bottom panel
    were measured by \citet{08FiZixx.VO}.
    }
    \label{fig:vostickcomp}

\end{figure*}

Figure\,\ref{fig:vostickcomp} illustrates the hyperfine splitting of 
non-hyperfine transitions
near \SI{9.77}{\per\cm}.
Due to the nuclear spin,
both the upper and lower non-hyperfine energy levels 
split to several hyperfine levels and
the combinations of them give a lot hyperfine transitions
as shown in the top panel.
In the middle panel,
we plot the two strongest non-hyperfine transition in this region.
The intensity of each non-hyperfine transition is
approximately the sum of intensities of
the eight strong hyperfine transitions nearby
but not rigorously equal to it.
These strong hyperfine transitions were observed by \citet{08FiZixx.VO}.
Our calculated positions agree well with 
the measured values.
Note that,
the hyperfine transitions are not necessarily
distributed around the non-hyperfine transitions,
as the transitions near \SI{9.8}{\per\cm} indicate.
We emphasize again that
in this paper the word `non-hyperfine' is used as shorthand notation
for the terms given without considering nuclear hyperfine interactions.
The word has a different meaning from `hyperfine unresolved'
which is used to describe blended hyperfine transitions.

As the nuclear spin of \ce{^{51}V^{16}O} is $7/2$,
theoretically, one can get `forbidden' dipole transitions up to $\abs{\Delta J} = 8$.
Table\,\ref{tab:vox_trans_s_branch} lists
eight transitions corresponding to $\abs{\Delta J} = 1, 2,  \cdots 8$.
As $J$ is no longer a good quantum number for hyperfine structure,
the $J'$ and $J''$ values here are the values of dominant basis functions.
The higher $\abs{\Delta J}$ transitions
are much weaker while 
transitions with $\abs{\Delta J}=2$ or $3$ have
has Einstein-$A$ of similar magnitude to
the `allowed' $\abs{\Delta J}=1$ one.
We are not aware of the observation of such forbidden lines within the $\VOXSigma$ state.
However,   $\abs{\Delta J}=2$ (O and S branches) 
driven by hyperfine couplings have been observed in both
hyperfine-resolved \cite{82ChHaMe,95AdBaBeBo} and unresolved \cite{94ChHaHu.VO,97KaLiLuSa} rovibronic
spectra.

\begin{table*}
\centering
\caption{Transitions corresponding to $\abs{\Delta J} = 1 \ldots 8$. }
\label{tab:vox_trans_s_branch}
\begin{ruledtabular}
\begin{tabular}{ccc|cccccc|cccccc}
$\abs{\Delta J}$    & $\nu$ [\si{\per\cm}]    & $A$  [\si{\per\second}]   & $E'$  [\si{\per\cm}]   & $F'$    & parity$'$ & $J'$    & $v'$    & $\varOmega'$ & $E''$ [\si{\per\cm}]   & $F''$   & parity$''$ & $J''$   & $v''$   & $\varOmega''$ \\
\hline
1 & 890.2463 & 6.7981E+01 & 9370.8835 & 0 & - & 3.5 & 10 & 0.5 & 8480.6371 & 1 & + & 4.5 & 9 & 0.5\\
2     & 880.4238 & 5.6402E+01 & 8635.9538 & 14    & +     & 16.5  & 9     & 1.5  & 7755.5299 & 15    & -     & 18.5  & 8     & 1.5 \\
3     & 921.1696 & 2.3566E+01 & 9975.2714 & 37    & -     & 35.5  & 10    & 0.5  & 9054.1018 & 36    & +     & 32.5  & 9     & 1.5 \\
4     & 942.0396 & 2.1003E-01 & 3909.3168 & 9     & -     & 6.5   & 4     & 0.5  & 2967.2771 & 10    & +     & 10.5  & 3     & 0.5 \\
5     & 925.0690 & 6.9014E-06 & 9437.0199 & 11    & +     & 11.5  & 10    & 1.5  & 8511.9509 & 10    & -     & 6.5   & 9     & 0.5 \\
6     & 923.5723 & 1.6667E-09 & 9435.5233 & 11    & +     & 12.5  & 10    & 0.5  & 8511.9509 & 10    & -     & 6.5   & 9     & 0.5 \\
7     & 950.5501 & 1.8763E-15 & 9462.5010 & 11    & +     & 13.5  & 10    & 1.5  & 8511.9509 & 10    & -     & 6.5   & 9     & 0.5 \\
8     & 949.0735 & 1.6573E-18 & 9461.0244 & 11    & +     & 14.5  & 10    & 0.5  & 8511.9509 & 10    & -     & 6.5   & 9     & 0.5 \\
\end{tabular}
\end{ruledtabular}
\end{table*}

\begin{figure*}
    \centering
    \includegraphics{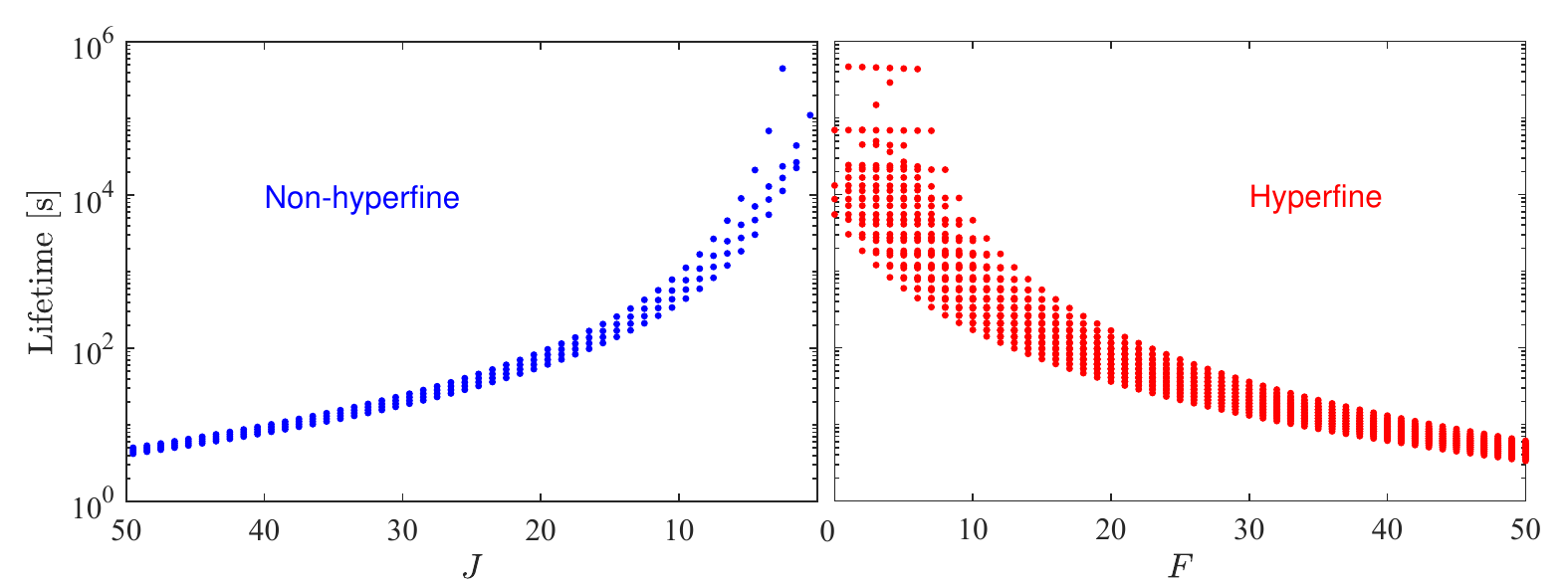}
    \caption{
    Comparison of
    lifetimes corresponding to
    the lower rotational levels of
    $\VOXSigma$, $v=0$.
    The $J=0.5$ levels 
    which have much longer lifetimes
    were not plotted in this figure.
    }
    \label{fig:volife}
\end{figure*}

The lifetimes of hyperfine
and non-hyperfine eigenstates
of the lowest vibrational level
of $\VOXSigma$ were
calculated by using \exocross,
and compared in Fig.\,\ref{fig:volife}.
The hyperfine states have similar lifetimes 
as the corresponding non-hyperfine state.

\section{Conclusion}

In this work,
we investigate the 
hyperfine-resolved infra-red spectra of VO $\mathrm{X}\,^4\Sigma^-$ electronic state.
The fine and hyperfine coupling 
curves required  to
construct the spectroscopic model were calculated \abinitio where possible but  then scaled to reproduce the observed hyperfine structure.
The hyperfine splitting of $\mathrm{X}\,^4\Sigma^-$  is mainly determined by
the Fermi-contact and
electron spin-nuclear spin dipolar interactions.
Nevertheless,
we also included the 
nuclear spin-rotation and
nuclear electric quadrupole coupling curves
in our calculation.
The hyperfine resolved and unresolved cross sections 
show good consistency with each other
when using wide line broadening parameters.
The comparison between calculated and
and empirical energy levels reveals 
the inaccuracy of our \abinitio fine and hyperfine coupling curves even when computed using state-of-the-art methods and hence the need for empirical refinement.
We plan to refine these curves and use them to
generate a full, hyperfine-resolved line list for VO in future work.

\section*{Supplementary Material}
The \duo input file used in this work 
is given as supplementary material;
our potential energy curves 
are included as part of this input file.
Two tables, 
which lists the sample states and 
transitions calculated from the input,
are given as supplementary materials.

\begin{acknowledgments}
We thank Gunnar Jeschke, Laura McKemmish 
and Charles Bowesman for valuable discussions. 
Qianwei Qu acknowledges the financial support from
University College London and China Scholarship Council.
This work was supported by the STFC Projects 
No. ST/M001334/1 and ST/R000476/1, 
and ERC Advanced Investigator Project 883830.
The authors acknowledge the use of the UCL Myriad
and Kathleen High Performance Computing
Facilities and 
associated support services in the completion of this work.
\end{acknowledgments}

\section*{Data Availability}

The open access  programs \exocross\ and \duo\ are  available from \href{https://github.com/exomol}{github.com/exomol}.

\bibliography{bib_journals_iso,bib_abinitio,bib_jtj,bib_methods,bib_programs,bib_VO,bib_VO_MARVEL,bib_TiO,bib_EPR,bib_hyperfine}

\begin{thebibliography}{47}%
\makeatletter
\providecommand \@ifxundefined [1]{%
 \@ifx{#1\undefined}
}%
\providecommand \@ifnum [1]{%
 \ifnum #1\expandafter \@firstoftwo
 \else \expandafter \@secondoftwo
 \fi
}%
\providecommand \@ifx [1]{%
 \ifx #1\expandafter \@firstoftwo
 \else \expandafter \@secondoftwo
 \fi
}%
\providecommand \natexlab [1]{#1}%
\providecommand \enquote  [1]{``#1''}%
\providecommand \bibnamefont  [1]{#1}%
\providecommand \bibfnamefont [1]{#1}%
\providecommand \citenamefont [1]{#1}%
\providecommand \href@noop [0]{\@secondoftwo}%
\providecommand \href [0]{\begingroup \@sanitize@url \@href}%
\providecommand \@href[1]{\@@startlink{#1}\@@href}%
\providecommand \@@href[1]{\endgroup#1\@@endlink}%
\providecommand \@sanitize@url [0]{\catcode `\\12\catcode `\$12\catcode
  `\&12\catcode `\#12\catcode `\^12\catcode `\_12\catcode `\%12\relax}%
\providecommand \@@startlink[1]{}%
\providecommand \@@endlink[0]{}%
\providecommand \url  [0]{\begingroup\@sanitize@url \@url }%
\providecommand \@url [1]{\endgroup\@href {#1}{\urlprefix }}%
\providecommand \urlprefix  [0]{URL }%
\providecommand \Eprint [0]{\href }%
\providecommand \doibase [0]{http://dx.doi.org/}%
\providecommand \selectlanguage [0]{\@gobble}%
\providecommand \bibinfo  [0]{\@secondoftwo}%
\providecommand \bibfield  [0]{\@secondoftwo}%
\providecommand \translation [1]{[#1]}%
\providecommand \BibitemOpen [0]{}%
\providecommand \bibitemStop [0]{}%
\providecommand \bibitemNoStop [0]{.\EOS\space}%
\providecommand \EOS [0]{\spacefactor3000\relax}%
\providecommand \BibitemShut  [1]{\csname bibitem#1\endcsname}%
\let\auto@bib@innerbib\@empty
%</preamble>
\bibitem [{\citenamefont {Bernath}(2009)}]{09Bernath.VO}%
  \BibitemOpen
  \bibfield  {author} {\bibinfo {author} {\bibfnamefont {P.~F.}\ \bibnamefont
  {Bernath}},\ }\href {\doibase 10.1080/01442350903292442} {\bibfield
  {journal} {\bibinfo  {journal} {Int. Rev. Phys. Chem.}\ }\textbf {\bibinfo
  {volume} {28}},\ \bibinfo {pages} {681} (\bibinfo {year} {2009})}\BibitemShut
  {NoStop}%
\bibitem [{\citenamefont {Madhusudhan}\ and\ \citenamefont
  {Seager}(2010)}]{10MaSexx.VO}%
  \BibitemOpen
  \bibfield  {author} {\bibinfo {author} {\bibfnamefont {N.}~\bibnamefont
  {Madhusudhan}}\ and\ \bibinfo {author} {\bibfnamefont {S.}~\bibnamefont
  {Seager}},\ }\href {\doibase 10.1088/0004-637X/725/1/261} {\bibfield
  {journal} {\bibinfo  {journal} {Astrophys. J.}\ }\textbf {\bibinfo {volume}
  {725}},\ \bibinfo {pages} {261} (\bibinfo {year} {2010})}\BibitemShut
  {NoStop}%
\bibitem [{\citenamefont {Evans}\ \emph {et~al.}(2016)\citenamefont {Evans},
  \citenamefont {Sing}, \citenamefont {Wakeford}, \citenamefont {Nikolov},
  \citenamefont {Ballester}, \citenamefont {Drummond}, \citenamefont {Kataria},
  \citenamefont {Gibson}, \citenamefont {Amundsen},\ and\ \citenamefont
  {Spake}}]{16EvSiWa.VO}%
  \BibitemOpen
  \bibfield  {author} {\bibinfo {author} {\bibfnamefont {T.~M.}\ \bibnamefont
  {Evans}}, \bibinfo {author} {\bibfnamefont {D.~K.}\ \bibnamefont {Sing}},
  \bibinfo {author} {\bibfnamefont {H.~R.}\ \bibnamefont {Wakeford}}, \bibinfo
  {author} {\bibfnamefont {N.}~\bibnamefont {Nikolov}}, \bibinfo {author}
  {\bibfnamefont {G.~E.}\ \bibnamefont {Ballester}}, \bibinfo {author}
  {\bibfnamefont {B.}~\bibnamefont {Drummond}}, \bibinfo {author}
  {\bibfnamefont {T.}~\bibnamefont {Kataria}}, \bibinfo {author} {\bibfnamefont
  {N.~P.}\ \bibnamefont {Gibson}}, \bibinfo {author} {\bibfnamefont {D.~S.}\
  \bibnamefont {Amundsen}}, \ and\ \bibinfo {author} {\bibfnamefont
  {J.}~\bibnamefont {Spake}},\ }\href {\doibase 10.3847/2041-8205/822/1/l4}
  {\bibfield  {journal} {\bibinfo  {journal} {Astrophys. J.}\ }\textbf
  {\bibinfo {volume} {822}},\ \bibinfo {pages} {L4} (\bibinfo {year}
  {2016})}\BibitemShut {NoStop}%
\bibitem [{\citenamefont {Turner}\ \emph {et~al.}(2017)\citenamefont {Turner},
  \citenamefont {Leiter}, \citenamefont {Biddle}, \citenamefont {Pearson},
  \citenamefont {Hardegree-Ullman}, \citenamefont {Thompson}, \citenamefont
  {Teske}, \citenamefont {Cates}, \citenamefont {Cook}, \citenamefont {Berube},
  \citenamefont {Nieberding}, \citenamefont {Jones}, \citenamefont {Raphael},
  \citenamefont {Wallace}, \citenamefont {Watson},\ and\ \citenamefont
  {Johnson}}]{17TuLeBi.VO}%
  \BibitemOpen
  \bibfield  {author} {\bibinfo {author} {\bibfnamefont {J.~D.}\ \bibnamefont
  {Turner}}, \bibinfo {author} {\bibfnamefont {R.~M.}\ \bibnamefont {Leiter}},
  \bibinfo {author} {\bibfnamefont {L.~I.}\ \bibnamefont {Biddle}}, \bibinfo
  {author} {\bibfnamefont {K.~A.}\ \bibnamefont {Pearson}}, \bibinfo {author}
  {\bibfnamefont {K.~K.}\ \bibnamefont {Hardegree-Ullman}}, \bibinfo {author}
  {\bibfnamefont {R.~M.}\ \bibnamefont {Thompson}}, \bibinfo {author}
  {\bibfnamefont {J.~K.}\ \bibnamefont {Teske}}, \bibinfo {author}
  {\bibfnamefont {I.~T.}\ \bibnamefont {Cates}}, \bibinfo {author}
  {\bibfnamefont {K.~L.}\ \bibnamefont {Cook}}, \bibinfo {author}
  {\bibfnamefont {M.~P.}\ \bibnamefont {Berube}}, \bibinfo {author}
  {\bibfnamefont {M.~N.}\ \bibnamefont {Nieberding}}, \bibinfo {author}
  {\bibfnamefont {C.~K.}\ \bibnamefont {Jones}}, \bibinfo {author}
  {\bibfnamefont {B.}~\bibnamefont {Raphael}}, \bibinfo {author} {\bibfnamefont
  {S.}~\bibnamefont {Wallace}}, \bibinfo {author} {\bibfnamefont {Z.~T.}\
  \bibnamefont {Watson}}, \ and\ \bibinfo {author} {\bibfnamefont {R.~E.}\
  \bibnamefont {Johnson}},\ }\href {\doibase 10.1093/mnras/stx2221} {\bibfield
  {journal} {\bibinfo  {journal} {Mon. Not. Roy. Astron. Soc.}\ }\textbf
  {\bibinfo {volume} {{472}}},\ \bibinfo {pages} {3871} (\bibinfo {year}
  {{2017}})}\BibitemShut {NoStop}%
\bibitem [{\citenamefont {Palle}\ \emph {et~al.}(2017)\citenamefont {Palle},
  \citenamefont {Chen}, \citenamefont {Prieto-Arranz}, \citenamefont {Nowak},
  \citenamefont {Murgas}, \citenamefont {Nortmann}, \citenamefont {Pollacco},
  \citenamefont {Lam}, \citenamefont {Montanes-Rodriguez}, \citenamefont
  {Parviainen},\ and\ \citenamefont {Casasayas-Barris}}]{17PaChPr.VO}%
  \BibitemOpen
  \bibfield  {author} {\bibinfo {author} {\bibfnamefont {E.}~\bibnamefont
  {Palle}}, \bibinfo {author} {\bibfnamefont {G.}~\bibnamefont {Chen}},
  \bibinfo {author} {\bibfnamefont {J.}~\bibnamefont {Prieto-Arranz}}, \bibinfo
  {author} {\bibfnamefont {G.}~\bibnamefont {Nowak}}, \bibinfo {author}
  {\bibfnamefont {F.}~\bibnamefont {Murgas}}, \bibinfo {author} {\bibfnamefont
  {L.}~\bibnamefont {Nortmann}}, \bibinfo {author} {\bibfnamefont
  {D.}~\bibnamefont {Pollacco}}, \bibinfo {author} {\bibfnamefont
  {K.}~\bibnamefont {Lam}}, \bibinfo {author} {\bibfnamefont {P.}~\bibnamefont
  {Montanes-Rodriguez}}, \bibinfo {author} {\bibfnamefont {H.}~\bibnamefont
  {Parviainen}}, \ and\ \bibinfo {author} {\bibfnamefont {N.}~\bibnamefont
  {Casasayas-Barris}},\ }\href {\doibase {10.1051/0004-6361/201731018}}
  {\bibfield  {journal} {\bibinfo  {journal} {Astron. Astrophys.}\ }\textbf
  {\bibinfo {volume} {{602}}} (\bibinfo {year} {{2017}}),\
  {10.1051/0004-6361/201731018}}\BibitemShut {NoStop}%
\bibitem [{\citenamefont {Tsiaras}\ \emph {et~al.}(2018)\citenamefont
  {Tsiaras}, \citenamefont {Waldmann}, \citenamefont {Zingales}, \citenamefont
  {Rocchetto}, \citenamefont {Morello}, \citenamefont {Damiano}, \citenamefont
  {Karpouzas}, \citenamefont {Tinetti}, \citenamefont {McKemmish},
  \citenamefont {Tennyson},\ and\ \citenamefont {Yurchenko}}]{jt699}%
  \BibitemOpen
  \bibfield  {author} {\bibinfo {author} {\bibfnamefont {A.}~\bibnamefont
  {Tsiaras}}, \bibinfo {author} {\bibfnamefont {I.~P.}\ \bibnamefont
  {Waldmann}}, \bibinfo {author} {\bibfnamefont {T.}~\bibnamefont {Zingales}},
  \bibinfo {author} {\bibfnamefont {M.}~\bibnamefont {Rocchetto}}, \bibinfo
  {author} {\bibfnamefont {G.}~\bibnamefont {Morello}}, \bibinfo {author}
  {\bibfnamefont {M.}~\bibnamefont {Damiano}}, \bibinfo {author} {\bibfnamefont
  {K.}~\bibnamefont {Karpouzas}}, \bibinfo {author} {\bibfnamefont
  {G.}~\bibnamefont {Tinetti}}, \bibinfo {author} {\bibfnamefont {L.~K.}\
  \bibnamefont {McKemmish}}, \bibinfo {author} {\bibfnamefont {J.}~\bibnamefont
  {Tennyson}}, \ and\ \bibinfo {author} {\bibfnamefont {S.~N.}\ \bibnamefont
  {Yurchenko}},\ }\href {\doibase 10.3847/1538-3881/aaaf75} {\bibfield
  {journal} {\bibinfo  {journal} {Astron. J.}\ }\textbf {\bibinfo {volume}
  {155}},\ \bibinfo {pages} {156} (\bibinfo {year} {2018})}\BibitemShut
  {NoStop}%
\bibitem [{\citenamefont {Goyal}\ \emph {et~al.}(2020)\citenamefont {Goyal},
  \citenamefont {Mayne}, \citenamefont {Drummond}, \citenamefont {Sing},
  \citenamefont {Hebrard}, \citenamefont {Lewis}, \citenamefont {Tremblin},
  \citenamefont {Phillips}, \citenamefont {Mikal-Evans},\ and\ \citenamefont
  {Wakeford}}]{20GoMaDrSi.VO}%
  \BibitemOpen
  \bibfield  {author} {\bibinfo {author} {\bibfnamefont {J.~M.}\ \bibnamefont
  {Goyal}}, \bibinfo {author} {\bibfnamefont {N.}~\bibnamefont {Mayne}},
  \bibinfo {author} {\bibfnamefont {B.}~\bibnamefont {Drummond}}, \bibinfo
  {author} {\bibfnamefont {D.~K.}\ \bibnamefont {Sing}}, \bibinfo {author}
  {\bibfnamefont {E.}~\bibnamefont {Hebrard}}, \bibinfo {author} {\bibfnamefont
  {N.}~\bibnamefont {Lewis}}, \bibinfo {author} {\bibfnamefont
  {P.}~\bibnamefont {Tremblin}}, \bibinfo {author} {\bibfnamefont {M.~W.}\
  \bibnamefont {Phillips}}, \bibinfo {author} {\bibfnamefont {T.}~\bibnamefont
  {Mikal-Evans}}, \ and\ \bibinfo {author} {\bibfnamefont {H.~R.}\ \bibnamefont
  {Wakeford}},\ }\href {\doibase 10.1093/mnras/staa2300} {\bibfield  {journal}
  {\bibinfo  {journal} {Mon. Not. Roy. Astron. Soc.}\ }\textbf {\bibinfo
  {volume} {498}},\ \bibinfo {pages} {4680} (\bibinfo {year}
  {2020})}\BibitemShut {NoStop}%
\bibitem [{\citenamefont {Lewis}\ \emph {et~al.}(2020)\citenamefont {Lewis},
  \citenamefont {Wakeford}, \citenamefont {MacDonald}, \citenamefont {Goyal},
  \citenamefont {Sing}, \citenamefont {Barstow}, \citenamefont {Powell},
  \citenamefont {Kataria}, \citenamefont {Mishra}, \citenamefont {Marley},
  \citenamefont {Batalha}, \citenamefont {Moses}, \citenamefont {Gao},
  \citenamefont {Wilson}, \citenamefont {Chubb}, \citenamefont {Mikal-Evans},
  \citenamefont {Nikolov}, \citenamefont {Pirzkal}, \citenamefont {Spake},
  \citenamefont {Stevenson}, \citenamefont {Valenti},\ and\ \citenamefont
  {Zhang}}]{20LeWaMa.VO}%
  \BibitemOpen
  \bibfield  {author} {\bibinfo {author} {\bibfnamefont {N.~K.}\ \bibnamefont
  {Lewis}}, \bibinfo {author} {\bibfnamefont {H.~R.}\ \bibnamefont {Wakeford}},
  \bibinfo {author} {\bibfnamefont {R.~J.}\ \bibnamefont {MacDonald}}, \bibinfo
  {author} {\bibfnamefont {J.~M.}\ \bibnamefont {Goyal}}, \bibinfo {author}
  {\bibfnamefont {D.~K.}\ \bibnamefont {Sing}}, \bibinfo {author}
  {\bibfnamefont {J.}~\bibnamefont {Barstow}}, \bibinfo {author} {\bibfnamefont
  {D.}~\bibnamefont {Powell}}, \bibinfo {author} {\bibfnamefont
  {T.}~\bibnamefont {Kataria}}, \bibinfo {author} {\bibfnamefont
  {I.}~\bibnamefont {Mishra}}, \bibinfo {author} {\bibfnamefont {M.~S.}\
  \bibnamefont {Marley}}, \bibinfo {author} {\bibfnamefont {N.~E.}\
  \bibnamefont {Batalha}}, \bibinfo {author} {\bibfnamefont {J.~I.}\
  \bibnamefont {Moses}}, \bibinfo {author} {\bibfnamefont {P.}~\bibnamefont
  {Gao}}, \bibinfo {author} {\bibfnamefont {T.~J.}\ \bibnamefont {Wilson}},
  \bibinfo {author} {\bibfnamefont {K.~L.}\ \bibnamefont {Chubb}}, \bibinfo
  {author} {\bibfnamefont {T.}~\bibnamefont {Mikal-Evans}}, \bibinfo {author}
  {\bibfnamefont {N.}~\bibnamefont {Nikolov}}, \bibinfo {author} {\bibfnamefont
  {N.}~\bibnamefont {Pirzkal}}, \bibinfo {author} {\bibfnamefont {J.~J.}\
  \bibnamefont {Spake}}, \bibinfo {author} {\bibfnamefont {K.~B.}\ \bibnamefont
  {Stevenson}}, \bibinfo {author} {\bibfnamefont {J.}~\bibnamefont {Valenti}},
  \ and\ \bibinfo {author} {\bibfnamefont {X.}~\bibnamefont {Zhang}},\ }\href
  {\doibase {10.3847/2041-8213/abb77f}} {\bibfield  {journal} {\bibinfo
  {journal} {Astrophys. J. Lett.}\ }\textbf {\bibinfo {volume} {{902}}},\
  \bibinfo {pages} {L19} (\bibinfo {year} {{2020}})}\BibitemShut {NoStop}%
\bibitem [{\citenamefont {McKemmish}\ \emph {et~al.}(2017)\citenamefont
  {McKemmish}, \citenamefont {Masseron}, \citenamefont {Sheppard},
  \citenamefont {Sandeman}, \citenamefont {Schofield}, \citenamefont
  {Furtenbacher}, \citenamefont {{Cs\'asz\'ar}}, \citenamefont {Tennyson},\
  and\ \citenamefont {Sousa-Silva}}]{jt672}%
  \BibitemOpen
  \bibfield  {author} {\bibinfo {author} {\bibfnamefont {L.~K.}\ \bibnamefont
  {McKemmish}}, \bibinfo {author} {\bibfnamefont {T.}~\bibnamefont {Masseron}},
  \bibinfo {author} {\bibfnamefont {S.}~\bibnamefont {Sheppard}}, \bibinfo
  {author} {\bibfnamefont {E.}~\bibnamefont {Sandeman}}, \bibinfo {author}
  {\bibfnamefont {Z.}~\bibnamefont {Schofield}}, \bibinfo {author}
  {\bibfnamefont {T.}~\bibnamefont {Furtenbacher}}, \bibinfo {author}
  {\bibfnamefont {A.~G.}\ \bibnamefont {{Cs\'asz\'ar}}}, \bibinfo {author}
  {\bibfnamefont {J.}~\bibnamefont {Tennyson}}, \ and\ \bibinfo {author}
  {\bibfnamefont {C.}~\bibnamefont {Sousa-Silva}},\ }\href {\doibase
  10.3847/1538-4365/228/2/15} {\bibfield  {journal} {\bibinfo  {journal}
  {Astrophys. J. Suppl. Ser.}\ }\textbf {\bibinfo {volume} {228}},\ \bibinfo
  {pages} {15} (\bibinfo {year} {2017})}\BibitemShut {NoStop}%
\bibitem [{\citenamefont {Nugroho}\ \emph {et~al.}(2017)\citenamefont
  {Nugroho}, \citenamefont {Kawahara}, \citenamefont {Masuda}, \citenamefont
  {Hirano}, \citenamefont {Kotani},\ and\ \citenamefont
  {Tajitsu}}]{17NuKaHa.TiO}%
  \BibitemOpen
  \bibfield  {author} {\bibinfo {author} {\bibfnamefont {S.~K.}\ \bibnamefont
  {Nugroho}}, \bibinfo {author} {\bibfnamefont {H.}~\bibnamefont {Kawahara}},
  \bibinfo {author} {\bibfnamefont {K.}~\bibnamefont {Masuda}}, \bibinfo
  {author} {\bibfnamefont {T.}~\bibnamefont {Hirano}}, \bibinfo {author}
  {\bibfnamefont {T.}~\bibnamefont {Kotani}}, \ and\ \bibinfo {author}
  {\bibfnamefont {A.}~\bibnamefont {Tajitsu}},\ }\href {\doibase
  10.3847/1538-3881/aa9433} {\bibfield  {journal} {\bibinfo  {journal}
  {Astrophys. J.}\ }\textbf {\bibinfo {volume} {154}},\ \bibinfo {pages} {221}
  (\bibinfo {year} {2017})}\BibitemShut {NoStop}%
\bibitem [{\citenamefont {Serindag}, \citenamefont {Snellen},\ and\
  \citenamefont {Molliere}(2021)}]{21SeSnMo.TiO}%
  \BibitemOpen
  \bibfield  {author} {\bibinfo {author} {\bibfnamefont {D.~B.}\ \bibnamefont
  {Serindag}}, \bibinfo {author} {\bibfnamefont {I.~A.~G.}\ \bibnamefont
  {Snellen}}, \ and\ \bibinfo {author} {\bibfnamefont {P.}~\bibnamefont
  {Molliere}},\ }\href {\doibase {10.1051/0004-6361/202141941}} {\bibfield
  {journal} {\bibinfo  {journal} {Astron. Astrophys.}\ }\textbf {\bibinfo
  {volume} {{655}}},\ \bibinfo {pages} {A69} (\bibinfo {year}
  {{2021}})}\BibitemShut {NoStop}%
\bibitem [{\citenamefont {Prinoth}\ \emph {et~al.}(2022)\citenamefont
  {Prinoth}, \citenamefont {Hoeijmakers}, \citenamefont {Kitzmann},
  \citenamefont {Sandvik}, \citenamefont {Seidel}, \citenamefont {Lendl},
  \citenamefont {Borsato}, \citenamefont {Thorsbro}, \citenamefont {Anderson},
  \citenamefont {Barrado}, \citenamefont {Kravchenko}, \citenamefont {Allart},
  \citenamefont {Bourrier}, \citenamefont {Cegla}, \citenamefont {Ehrenreich},
  \citenamefont {Fisher}, \citenamefont {Lovis}, \citenamefont {Guzman-Mesa},
  \citenamefont {Grimm}, \citenamefont {Hooton}, \citenamefont {Morris},
  \citenamefont {Oreshenko}, \citenamefont {Pino},\ and\ \citenamefont
  {Heng}}]{22BiHoKi.TiO}%
  \BibitemOpen
  \bibfield  {author} {\bibinfo {author} {\bibfnamefont {B.}~\bibnamefont
  {Prinoth}}, \bibinfo {author} {\bibfnamefont {H.~J.}\ \bibnamefont
  {Hoeijmakers}}, \bibinfo {author} {\bibfnamefont {D.}~\bibnamefont
  {Kitzmann}}, \bibinfo {author} {\bibfnamefont {E.}~\bibnamefont {Sandvik}},
  \bibinfo {author} {\bibfnamefont {J.}~\bibnamefont {Seidel}, \bibfnamefont
  {V}}, \bibinfo {author} {\bibfnamefont {M.}~\bibnamefont {Lendl}}, \bibinfo
  {author} {\bibfnamefont {N.~W.}\ \bibnamefont {Borsato}}, \bibinfo {author}
  {\bibfnamefont {B.}~\bibnamefont {Thorsbro}}, \bibinfo {author}
  {\bibfnamefont {D.~R.}\ \bibnamefont {Anderson}}, \bibinfo {author}
  {\bibfnamefont {D.}~\bibnamefont {Barrado}}, \bibinfo {author} {\bibfnamefont
  {K.}~\bibnamefont {Kravchenko}}, \bibinfo {author} {\bibfnamefont
  {R.}~\bibnamefont {Allart}}, \bibinfo {author} {\bibfnamefont
  {V.}~\bibnamefont {Bourrier}}, \bibinfo {author} {\bibfnamefont {H.~M.}\
  \bibnamefont {Cegla}}, \bibinfo {author} {\bibfnamefont {D.}~\bibnamefont
  {Ehrenreich}}, \bibinfo {author} {\bibfnamefont {C.}~\bibnamefont {Fisher}},
  \bibinfo {author} {\bibfnamefont {C.}~\bibnamefont {Lovis}}, \bibinfo
  {author} {\bibfnamefont {A.}~\bibnamefont {Guzman-Mesa}}, \bibinfo {author}
  {\bibfnamefont {S.}~\bibnamefont {Grimm}}, \bibinfo {author} {\bibfnamefont
  {M.}~\bibnamefont {Hooton}}, \bibinfo {author} {\bibfnamefont {B.~M.}\
  \bibnamefont {Morris}}, \bibinfo {author} {\bibfnamefont {M.}~\bibnamefont
  {Oreshenko}}, \bibinfo {author} {\bibfnamefont {L.}~\bibnamefont {Pino}}, \
  and\ \bibinfo {author} {\bibfnamefont {K.}~\bibnamefont {Heng}},\ }\href
  {\doibase {10.1038/s41550-021-01581-z}} {\bibfield  {journal} {\bibinfo
  {journal} {Nature Astr.}\ }\textbf {\bibinfo {volume} {6}},\ \bibinfo {pages}
  {449} (\bibinfo {year} {2022})}\BibitemShut {NoStop}%
\bibitem [{\citenamefont {McKemmish}, \citenamefont {Yurchenko},\ and\
  \citenamefont {Tennyson}(2016{\natexlab{a}})}]{jt644}%
  \BibitemOpen
  \bibfield  {author} {\bibinfo {author} {\bibfnamefont {L.~K.}\ \bibnamefont
  {McKemmish}}, \bibinfo {author} {\bibfnamefont {S.~N.}\ \bibnamefont
  {Yurchenko}}, \ and\ \bibinfo {author} {\bibfnamefont {J.}~\bibnamefont
  {Tennyson}},\ }\href {\doibase 10.1093/mnras/stw1969} {\bibfield  {journal}
  {\bibinfo  {journal} {Mon. Not. Roy. Astron. Soc.}\ }\textbf {\bibinfo
  {volume} {463}},\ \bibinfo {pages} {771} (\bibinfo {year}
  {2016}{\natexlab{a}})}\BibitemShut {NoStop}%
\bibitem [{\citenamefont {{de Regt}}\ \emph {et~al.}(2022)\citenamefont {{de
  Regt}}, \citenamefont {Kesseli}, \citenamefont {Snellen}, \citenamefont
  {Merritt},\ and\ \citenamefont {Chubb}}]{22DeKeSn.VO}%
  \BibitemOpen
  \bibfield  {author} {\bibinfo {author} {\bibfnamefont {S.}~\bibnamefont {{de
  Regt}}}, \bibinfo {author} {\bibfnamefont {A.~Y.}\ \bibnamefont {Kesseli}},
  \bibinfo {author} {\bibfnamefont {I.~A.~G.}\ \bibnamefont {Snellen}},
  \bibinfo {author} {\bibfnamefont {S.~R.}\ \bibnamefont {Merritt}}, \ and\
  \bibinfo {author} {\bibfnamefont {K.~L.}\ \bibnamefont {Chubb}},\ }\href
  {\doibase 10.1051/0004-6361/202142683} {\bibfield  {journal} {\bibinfo
  {journal} {Astron. Astrophys.}\ }\textbf {\bibinfo {volume} {661}},\ \bibinfo
  {pages} {A109} (\bibinfo {year} {2022})}\BibitemShut {NoStop}%
\bibitem [{\citenamefont {Tennyson}\ \emph {et~al.}(2020)\citenamefont
  {Tennyson}, \citenamefont {Yurchenko}, \citenamefont {Al-Refaie},
  \citenamefont {Clark}, \citenamefont {Chubb}, \citenamefont {Conway},
  \citenamefont {Dewan}, \citenamefont {Gorman}, \citenamefont {Hill},
  \citenamefont {Lynas-Gray}, \citenamefont {Mellor}, \citenamefont
  {McKemmish}, \citenamefont {Owens}, \citenamefont {Polyansky}, \citenamefont
  {Semenov}, \citenamefont {Somogyi}, \citenamefont {Tinetti}, \citenamefont
  {Upadhyay}, \citenamefont {Waldmann}, \citenamefont {Wang}, \citenamefont
  {Wright},\ and\ \citenamefont {Yurchenko}}]{jt810}%
  \BibitemOpen
  \bibfield  {author} {\bibinfo {author} {\bibfnamefont {J.}~\bibnamefont
  {Tennyson}}, \bibinfo {author} {\bibfnamefont {S.~N.}\ \bibnamefont
  {Yurchenko}}, \bibinfo {author} {\bibfnamefont {A.~F.}\ \bibnamefont
  {Al-Refaie}}, \bibinfo {author} {\bibfnamefont {V.~H.~J.}\ \bibnamefont
  {Clark}}, \bibinfo {author} {\bibfnamefont {K.~L.}\ \bibnamefont {Chubb}},
  \bibinfo {author} {\bibfnamefont {E.~K.}\ \bibnamefont {Conway}}, \bibinfo
  {author} {\bibfnamefont {A.}~\bibnamefont {Dewan}}, \bibinfo {author}
  {\bibfnamefont {M.~N.}\ \bibnamefont {Gorman}}, \bibinfo {author}
  {\bibfnamefont {C.}~\bibnamefont {Hill}}, \bibinfo {author} {\bibfnamefont
  {A.~E.}\ \bibnamefont {Lynas-Gray}}, \bibinfo {author} {\bibfnamefont
  {T.}~\bibnamefont {Mellor}}, \bibinfo {author} {\bibfnamefont {L.~K.}\
  \bibnamefont {McKemmish}}, \bibinfo {author} {\bibfnamefont {A.}~\bibnamefont
  {Owens}}, \bibinfo {author} {\bibfnamefont {O.~L.}\ \bibnamefont
  {Polyansky}}, \bibinfo {author} {\bibfnamefont {M.}~\bibnamefont {Semenov}},
  \bibinfo {author} {\bibfnamefont {W.}~\bibnamefont {Somogyi}}, \bibinfo
  {author} {\bibfnamefont {G.}~\bibnamefont {Tinetti}}, \bibinfo {author}
  {\bibfnamefont {A.}~\bibnamefont {Upadhyay}}, \bibinfo {author}
  {\bibfnamefont {I.}~\bibnamefont {Waldmann}}, \bibinfo {author}
  {\bibfnamefont {Y.}~\bibnamefont {Wang}}, \bibinfo {author} {\bibfnamefont
  {S.}~\bibnamefont {Wright}}, \ and\ \bibinfo {author} {\bibfnamefont {O.~P.}\
  \bibnamefont {Yurchenko}},\ }\href {\doibase 10.1016/j.jqsrt.2020.107228}
  {\bibfield  {journal} {\bibinfo  {journal} {J. Quant. Spectrosc. Radiat.
  Transf.}\ }\textbf {\bibinfo {volume} {255}},\ \bibinfo {pages} {107228}
  (\bibinfo {year} {2020})}\BibitemShut {NoStop}%
\bibitem [{\citenamefont {Merer}(1989)}]{89Merer.VO}%
  \BibitemOpen
  \bibfield  {author} {\bibinfo {author} {\bibfnamefont {A.~J.}\ \bibnamefont
  {Merer}},\ }\href {\doibase 10.1146/annurev.pc.40.100189.002203} {\bibfield
  {journal} {\bibinfo  {journal} {Annu. Rev. Phys. Chem.}\ }\textbf {\bibinfo
  {volume} {40}},\ \bibinfo {pages} {407} (\bibinfo {year} {1989})}\BibitemShut
  {NoStop}%
\bibitem [{\citenamefont {Bowesman}\ \emph {et~al.}(2022)\citenamefont
  {Bowesman}, \citenamefont {Akbari}, \citenamefont {Hopkins}, \citenamefont
  {Yurchenko},\ and\ \citenamefont {Tennyson}}]{jt869}%
  \BibitemOpen
  \bibfield  {author} {\bibinfo {author} {\bibfnamefont {C.~A.}\ \bibnamefont
  {Bowesman}}, \bibinfo {author} {\bibfnamefont {H.}~\bibnamefont {Akbari}},
  \bibinfo {author} {\bibfnamefont {S.}~\bibnamefont {Hopkins}}, \bibinfo
  {author} {\bibfnamefont {S.~N.}\ \bibnamefont {Yurchenko}}, \ and\ \bibinfo
  {author} {\bibfnamefont {J.}~\bibnamefont {Tennyson}},\ }\href {\doibase
  10.1016/j.jqsrt.2022.108295} {\bibfield  {journal} {\bibinfo  {journal} {J.
  Quant. Spectrosc. Radiat. Transf.}\ } (\bibinfo {year} {2022}),\
  10.1016/j.jqsrt.2022.108295}\BibitemShut {NoStop}%
\bibitem [{\citenamefont {Hocking}, \citenamefont {Merer},\ and\ \citenamefont
  {Milton}(1981)}]{81HoMeMi}%
  \BibitemOpen
  \bibfield  {author} {\bibinfo {author} {\bibfnamefont {W.~H.}\ \bibnamefont
  {Hocking}}, \bibinfo {author} {\bibfnamefont {A.~J.}\ \bibnamefont {Merer}},
  \ and\ \bibinfo {author} {\bibfnamefont {D.~J.}\ \bibnamefont {Milton}},\
  }\href {\doibase 10.1139/p81-035} {\bibfield  {journal} {\bibinfo  {journal}
  {Can. J. Phys.}\ }\textbf {\bibinfo {volume} {59}},\ \bibinfo {pages} {266}
  (\bibinfo {year} {1981})}\BibitemShut {NoStop}%
\bibitem [{\citenamefont {Cheung}, \citenamefont {Hansen},\ and\ \citenamefont
  {Merer}(1982)}]{82ChHaMe}%
  \BibitemOpen
  \bibfield  {author} {\bibinfo {author} {\bibfnamefont {A.~S.-C.}\
  \bibnamefont {Cheung}}, \bibinfo {author} {\bibfnamefont {R.~C.}\
  \bibnamefont {Hansen}}, \ and\ \bibinfo {author} {\bibfnamefont {A.~J.}\
  \bibnamefont {Merer}},\ }\href {\doibase 10.1016/0022-2852(82)90039-X}
  {\bibfield  {journal} {\bibinfo  {journal} {J. Mol. Spectrosc.}\ }\textbf
  {\bibinfo {volume} {91}},\ \bibinfo {pages} {165} (\bibinfo {year}
  {1982})}\BibitemShut {NoStop}%
\bibitem [{\citenamefont {Suenram}\ \emph {et~al.}(1991)\citenamefont
  {Suenram}, \citenamefont {Fraser}, \citenamefont {Lovas},\ and\ \citenamefont
  {Gillies}}]{91SuFrLoGi}%
  \BibitemOpen
  \bibfield  {author} {\bibinfo {author} {\bibfnamefont {R.~D.}\ \bibnamefont
  {Suenram}}, \bibinfo {author} {\bibfnamefont {G.~T.}\ \bibnamefont {Fraser}},
  \bibinfo {author} {\bibfnamefont {F.~J.}\ \bibnamefont {Lovas}}, \ and\
  \bibinfo {author} {\bibfnamefont {C.~W.}\ \bibnamefont {Gillies}},\ }\href
  {\doibase 10.1016/0022-2852(91)90040-H} {\bibfield  {journal} {\bibinfo
  {journal} {J. Mol. Spectrosc.}\ }\textbf {\bibinfo {volume} {148}},\ \bibinfo
  {pages} {114} (\bibinfo {year} {1991})}\BibitemShut {NoStop}%
\bibitem [{\citenamefont {Adam}\ \emph {et~al.}(1995)\citenamefont {Adam},
  \citenamefont {Barnes}, \citenamefont {Berno}, \citenamefont {Bower},\ and\
  \citenamefont {Merer}}]{95AdBaBeBo}%
  \BibitemOpen
  \bibfield  {author} {\bibinfo {author} {\bibfnamefont {A.~G.}\ \bibnamefont
  {Adam}}, \bibinfo {author} {\bibfnamefont {M.}~\bibnamefont {Barnes}},
  \bibinfo {author} {\bibfnamefont {B.}~\bibnamefont {Berno}}, \bibinfo
  {author} {\bibfnamefont {R.~D.}\ \bibnamefont {Bower}}, \ and\ \bibinfo
  {author} {\bibfnamefont {A.~J.}\ \bibnamefont {Merer}},\ }\href {\doibase
  10.1006/jmsp.1995.1059} {\bibfield  {journal} {\bibinfo  {journal} {J. Mol.
  Spectrosc.}\ }\textbf {\bibinfo {volume} {170}},\ \bibinfo {pages} {94}
  (\bibinfo {year} {1995})}\BibitemShut {NoStop}%
\bibitem [{\citenamefont {Flory}\ and\ \citenamefont
  {Ziurys}(2008)}]{08FiZixx.VO}%
  \BibitemOpen
  \bibfield  {author} {\bibinfo {author} {\bibfnamefont {M.~A.}\ \bibnamefont
  {Flory}}\ and\ \bibinfo {author} {\bibfnamefont {L.~M.}\ \bibnamefont
  {Ziurys}},\ }\href {\doibase 10.1016/j.jms.2007.09.007} {\bibfield  {journal}
  {\bibinfo  {journal} {J. Mol. Spectrosc.}\ }\textbf {\bibinfo {volume}
  {247}},\ \bibinfo {pages} {76} (\bibinfo {year} {2008})}\BibitemShut
  {NoStop}%
\bibitem [{\citenamefont {Cheung}\ \emph {et~al.}(1994)\citenamefont {Cheung},
  \citenamefont {Hajigeorgiou}, \citenamefont {Huang}, \citenamefont {Huang},\
  and\ \citenamefont {Merer}}]{94ChHaHu.VO}%
  \BibitemOpen
  \bibfield  {author} {\bibinfo {author} {\bibfnamefont {A.~S.~C.}\
  \bibnamefont {Cheung}}, \bibinfo {author} {\bibfnamefont {P.~G.}\
  \bibnamefont {Hajigeorgiou}}, \bibinfo {author} {\bibfnamefont
  {G.}~\bibnamefont {Huang}}, \bibinfo {author} {\bibfnamefont {S.~Z.}\
  \bibnamefont {Huang}}, \ and\ \bibinfo {author} {\bibfnamefont {A.~J.}\
  \bibnamefont {Merer}},\ }\href {\doibase 10.1006/jmsp.1994.1039} {\bibfield
  {journal} {\bibinfo  {journal} {J. Mol. Spectrosc.}\ }\textbf {\bibinfo
  {volume} {163}},\ \bibinfo {pages} {443} (\bibinfo {year}
  {1994})}\BibitemShut {NoStop}%
\bibitem [{\citenamefont {Yurchenko}\ \emph {et~al.}(2016)\citenamefont
  {Yurchenko}, \citenamefont {Lodi}, \citenamefont {Tennyson},\ and\
  \citenamefont {Stolyarov}}]{jt609}%
  \BibitemOpen
  \bibfield  {author} {\bibinfo {author} {\bibfnamefont {S.~N.}\ \bibnamefont
  {Yurchenko}}, \bibinfo {author} {\bibfnamefont {L.}~\bibnamefont {Lodi}},
  \bibinfo {author} {\bibfnamefont {J.}~\bibnamefont {Tennyson}}, \ and\
  \bibinfo {author} {\bibfnamefont {A.~V.}\ \bibnamefont {Stolyarov}},\ }\href
  {\doibase 10.1016/j.cpc.2015.12.021} {\bibfield  {journal} {\bibinfo
  {journal} {Comput. Phys. Commun.}\ }\textbf {\bibinfo {volume} {202}},\
  \bibinfo {pages} {262} (\bibinfo {year} {2016})}\BibitemShut {NoStop}%
\bibitem [{\citenamefont {Qu}, \citenamefont {Yurchenko},\ and\ \citenamefont
  {Tennyson}(2022)}]{jt855}%
  \BibitemOpen
  \bibfield  {author} {\bibinfo {author} {\bibfnamefont {Q.}~\bibnamefont
  {Qu}}, \bibinfo {author} {\bibfnamefont {S.~N.}\ \bibnamefont {Yurchenko}}, \
  and\ \bibinfo {author} {\bibfnamefont {J.}~\bibnamefont {Tennyson}},\ }\href
  {\doibase 10.1021/acs.jctc.1c01244} {\bibfield  {journal} {\bibinfo
  {journal} {J. Chem. Theory Comput.}\ }\textbf {\bibinfo {volume} {18}},\
  \bibinfo {pages} {1808} (\bibinfo {year} {2022})}\BibitemShut {NoStop}%
\bibitem [{\citenamefont {Bauschlicher}\ and\ \citenamefont
  {Maitre}(1995)}]{95BaMaxx.VO}%
  \BibitemOpen
  \bibfield  {author} {\bibinfo {author} {\bibfnamefont {C.~W.}\ \bibnamefont
  {Bauschlicher}}\ and\ \bibinfo {author} {\bibfnamefont {P.}~\bibnamefont
  {Maitre}},\ }\href {\doibase 10.1007/BF01113847} {\bibfield  {journal}
  {\bibinfo  {journal} {Theor. Chim. Acta.}\ }\textbf {\bibinfo {volume}
  {90}},\ \bibinfo {pages} {189} (\bibinfo {year} {1995})}\BibitemShut
  {NoStop}%
\bibitem [{\citenamefont {Bridgeman}\ and\ \citenamefont
  {Rothery}(2000)}]{00BrRoxx.VO}%
  \BibitemOpen
  \bibfield  {author} {\bibinfo {author} {\bibfnamefont {A.~J.}\ \bibnamefont
  {Bridgeman}}\ and\ \bibinfo {author} {\bibfnamefont {J.}~\bibnamefont
  {Rothery}},\ }\href {\doibase 10.1039/a906523g} {\bibfield  {journal}
  {\bibinfo  {journal} {J. Chem. Soc. Dalton}\ ,\ \bibinfo {pages} {211}}
  (\bibinfo {year} {2000})}\BibitemShut {NoStop}%
\bibitem [{\citenamefont {Calatayud}\ \emph {et~al.}(2001)\citenamefont
  {Calatayud}, \citenamefont {Silvi}, \citenamefont {Andres},\ and\
  \citenamefont {Beltran}}]{01CaSiAn.VO}%
  \BibitemOpen
  \bibfield  {author} {\bibinfo {author} {\bibfnamefont {M.}~\bibnamefont
  {Calatayud}}, \bibinfo {author} {\bibfnamefont {B.}~\bibnamefont {Silvi}},
  \bibinfo {author} {\bibfnamefont {J.}~\bibnamefont {Andres}}, \ and\ \bibinfo
  {author} {\bibfnamefont {A.}~\bibnamefont {Beltran}},\ }\href {\doibase
  10.1016/S0009-2614(00)01287-2} {\bibfield  {journal} {\bibinfo  {journal}
  {Chem. Phys. Lett.}\ }\textbf {\bibinfo {volume} {333}},\ \bibinfo {pages}
  {493} (\bibinfo {year} {2001})}\BibitemShut {NoStop}%
\bibitem [{\citenamefont {Broclawik}\ and\ \citenamefont
  {Borowski}(2001)}]{01BrBoxx.VO}%
  \BibitemOpen
  \bibfield  {author} {\bibinfo {author} {\bibfnamefont {E.}~\bibnamefont
  {Broclawik}}\ and\ \bibinfo {author} {\bibfnamefont {T.}~\bibnamefont
  {Borowski}},\ }\href {\doibase 10.1016/S0009-2614(01)00361-X} {\bibfield
  {journal} {\bibinfo  {journal} {Chem. Phys. Lett.}\ }\textbf {\bibinfo
  {volume} {339}},\ \bibinfo {pages} {433} (\bibinfo {year}
  {2001})}\BibitemShut {NoStop}%
\bibitem [{\citenamefont {Dai}\ \emph {et~al.}(2003)\citenamefont {Dai},
  \citenamefont {Deng}, \citenamefont {Yang},\ and\ \citenamefont
  {Zhu}}]{03DaDeYa.VO}%
  \BibitemOpen
  \bibfield  {author} {\bibinfo {author} {\bibfnamefont {B.}~\bibnamefont
  {Dai}}, \bibinfo {author} {\bibfnamefont {K.~M.}\ \bibnamefont {Deng}},
  \bibinfo {author} {\bibfnamefont {J.~L.}\ \bibnamefont {Yang}}, \ and\
  \bibinfo {author} {\bibfnamefont {Q.~S.}\ \bibnamefont {Zhu}},\ }\href
  {\doibase 10.1063/1.1570811} {\bibfield  {journal} {\bibinfo  {journal} {J.
  Chem. Phys.}\ }\textbf {\bibinfo {volume} {118}},\ \bibinfo {pages} {9608}
  (\bibinfo {year} {2003})}\BibitemShut {NoStop}%
\bibitem [{\citenamefont {Pykavy}\ and\ \citenamefont {van
  Wullen}(2003)}]{03Pyvaxx.VO}%
  \BibitemOpen
  \bibfield  {author} {\bibinfo {author} {\bibfnamefont {M.}~\bibnamefont
  {Pykavy}}\ and\ \bibinfo {author} {\bibfnamefont {C.}~\bibnamefont {van
  Wullen}},\ }\href {\doibase 10.1021/jp027264n} {\bibfield  {journal}
  {\bibinfo  {journal} {J. Phys. Chem. A}\ }\textbf {\bibinfo {volume} {107}},\
  \bibinfo {pages} {5566} (\bibinfo {year} {2003})}\BibitemShut {NoStop}%
\bibitem [{\citenamefont {Kulik}\ and\ \citenamefont
  {Marzari}(2010)}]{10KuMaxx.VO}%
  \BibitemOpen
  \bibfield  {author} {\bibinfo {author} {\bibfnamefont {H.~J.}\ \bibnamefont
  {Kulik}}\ and\ \bibinfo {author} {\bibfnamefont {N.}~\bibnamefont
  {Marzari}},\ }\href {\doibase 10.1063/1.3489110} {\bibfield  {journal}
  {\bibinfo  {journal} {J. Chem. Phys.}\ }\textbf {\bibinfo {volume} {133}},\
  \bibinfo {pages} {114103} (\bibinfo {year} {2010})}\BibitemShut {NoStop}%
\bibitem [{\citenamefont {Miliordos}\ and\ \citenamefont
  {Mavridis}(2007)}]{07MiMaxx.VO}%
  \BibitemOpen
  \bibfield  {author} {\bibinfo {author} {\bibfnamefont {E.}~\bibnamefont
  {Miliordos}}\ and\ \bibinfo {author} {\bibfnamefont {A.}~\bibnamefont
  {Mavridis}},\ }\href {\doibase 10.1021/jp067451b} {\bibfield  {journal}
  {\bibinfo  {journal} {J. Phys. Chem. A}\ }\textbf {\bibinfo {volume} {111}},\
  \bibinfo {pages} {1953} (\bibinfo {year} {2007})}\BibitemShut {NoStop}%
\bibitem [{\citenamefont {H{\"u}bner}, \citenamefont {Hornung},\ and\
  \citenamefont {Himmel}(2015)}]{15HuHoHi.VO}%
  \BibitemOpen
  \bibfield  {author} {\bibinfo {author} {\bibfnamefont {O.}~\bibnamefont
  {H{\"u}bner}}, \bibinfo {author} {\bibfnamefont {J.}~\bibnamefont {Hornung}},
  \ and\ \bibinfo {author} {\bibfnamefont {H.-J.}\ \bibnamefont {Himmel}},\
  }\href {\doibase 10.1063/1.4926393} {\bibfield  {journal} {\bibinfo
  {journal} {J. Chem. Phys.}\ }\textbf {\bibinfo {volume} {143}},\ \bibinfo
  {pages} {024309} (\bibinfo {year} {2015})}\BibitemShut {NoStop}%
\bibitem [{\citenamefont {McKemmish}, \citenamefont {Yurchenko},\ and\
  \citenamefont {Tennyson}(2016{\natexlab{b}})}]{jt623}%
  \BibitemOpen
  \bibfield  {author} {\bibinfo {author} {\bibfnamefont {L.~K.}\ \bibnamefont
  {McKemmish}}, \bibinfo {author} {\bibfnamefont {S.~N.}\ \bibnamefont
  {Yurchenko}}, \ and\ \bibinfo {author} {\bibfnamefont {J.}~\bibnamefont
  {Tennyson}},\ }\href {\doibase 10.1080/00268976.2016.1225994} {\bibfield
  {journal} {\bibinfo  {journal} {Mol. Phys.}\ }\textbf {\bibinfo {volume}
  {114}},\ \bibinfo {pages} {3232} (\bibinfo {year}
  {2016}{\natexlab{b}})}\BibitemShut {NoStop}%
\bibitem [{\citenamefont {Jiang}\ \emph {et~al.}(2021)\citenamefont {Jiang},
  \citenamefont {Chen}, \citenamefont {Bogdanov}, \citenamefont {Wang},
  \citenamefont {Alavi},\ and\ \citenamefont {Chen}}]{21JiChBo.VO}%
  \BibitemOpen
  \bibfield  {author} {\bibinfo {author} {\bibfnamefont {T.}~\bibnamefont
  {Jiang}}, \bibinfo {author} {\bibfnamefont {Y.}~\bibnamefont {Chen}},
  \bibinfo {author} {\bibfnamefont {N.~A.}\ \bibnamefont {Bogdanov}}, \bibinfo
  {author} {\bibfnamefont {E.}~\bibnamefont {Wang}}, \bibinfo {author}
  {\bibfnamefont {A.}~\bibnamefont {Alavi}}, \ and\ \bibinfo {author}
  {\bibfnamefont {J.}~\bibnamefont {Chen}},\ }\href {\doibase
  10.1063/5.0046464} {\bibfield  {journal} {\bibinfo  {journal} {J. Chem.
  Phys.}\ ,\ \bibinfo {pages} {164302}} (\bibinfo {year} {2021})}\BibitemShut
  {NoStop}%
\bibitem [{\citenamefont {Werner}\ \emph {et~al.}(2015)\citenamefont {Werner},
  \citenamefont {Knowles}, \citenamefont {Knizia}, \citenamefont {Manby},
  \citenamefont {Sch{\"{u}}tz}, \citenamefont {Celani}, \citenamefont
  {Gy{\"{o}}rffy}, \citenamefont {Kats}, \citenamefont {Korona}, \citenamefont
  {Lindh}, \citenamefont {Mitrushenkov}, \citenamefont {Rauhut}, \citenamefont
  {Shamasundar}, \citenamefont {Adler}, \citenamefont {Amos}, \citenamefont
  {Bernhardsson}, \citenamefont {Berning}, \citenamefont {Cooper},
  \citenamefont {Deegan}, \citenamefont {Dobbyn}, \citenamefont {Eckert},
  \citenamefont {Goll}, \citenamefont {Hampel}, \citenamefont {Hesselmann},
  \citenamefont {Hetzer}, \citenamefont {Hrenar}, \citenamefont {Jansen},
  \citenamefont {K{\"{o}}ppl}, \citenamefont {Liu}, \citenamefont {Lloyd},
  \citenamefont {Mata}, \citenamefont {May}, \citenamefont {McNicholas},
  \citenamefont {Meyer}, \citenamefont {Mura}, \citenamefont {Nicklass},
  \citenamefont {O'Neill}, \citenamefont {Palmieri}, \citenamefont {Peng},
  \citenamefont {Pfl{\"{u}}ger}, \citenamefont {Pitzer}, \citenamefont
  {Reiher}, \citenamefont {Shiozaki}, \citenamefont {Stoll}, \citenamefont
  {Stone}, \citenamefont {Tarroni}, \citenamefont {Thorsteinsson},\ and\
  \citenamefont {Wang}}]{MOLPRO2015}%
  \BibitemOpen
  \bibfield  {author} {\bibinfo {author} {\bibfnamefont {H.~J.}\ \bibnamefont
  {Werner}}, \bibinfo {author} {\bibfnamefont {P.~J.}\ \bibnamefont {Knowles}},
  \bibinfo {author} {\bibfnamefont {G.}~\bibnamefont {Knizia}}, \bibinfo
  {author} {\bibfnamefont {F.~R.}\ \bibnamefont {Manby}}, \bibinfo {author}
  {\bibfnamefont {M.}~\bibnamefont {Sch{\"{u}}tz}}, \bibinfo {author}
  {\bibfnamefont {P.}~\bibnamefont {Celani}}, \bibinfo {author} {\bibfnamefont
  {W.}~\bibnamefont {Gy{\"{o}}rffy}}, \bibinfo {author} {\bibfnamefont
  {D.}~\bibnamefont {Kats}}, \bibinfo {author} {\bibfnamefont {T.}~\bibnamefont
  {Korona}}, \bibinfo {author} {\bibfnamefont {R.}~\bibnamefont {Lindh}},
  \bibinfo {author} {\bibfnamefont {A.}~\bibnamefont {Mitrushenkov}}, \bibinfo
  {author} {\bibfnamefont {G.}~\bibnamefont {Rauhut}}, \bibinfo {author}
  {\bibfnamefont {K.~R.}\ \bibnamefont {Shamasundar}}, \bibinfo {author}
  {\bibfnamefont {T.~B.}\ \bibnamefont {Adler}}, \bibinfo {author}
  {\bibfnamefont {R.~D.}\ \bibnamefont {Amos}}, \bibinfo {author}
  {\bibfnamefont {A.}~\bibnamefont {Bernhardsson}}, \bibinfo {author}
  {\bibfnamefont {A.}~\bibnamefont {Berning}}, \bibinfo {author} {\bibfnamefont
  {D.~L.}\ \bibnamefont {Cooper}}, \bibinfo {author} {\bibfnamefont {M.~J.~O.}\
  \bibnamefont {Deegan}}, \bibinfo {author} {\bibfnamefont {A.~J.}\
  \bibnamefont {Dobbyn}}, \bibinfo {author} {\bibfnamefont {F.}~\bibnamefont
  {Eckert}}, \bibinfo {author} {\bibfnamefont {E.}~\bibnamefont {Goll}},
  \bibinfo {author} {\bibfnamefont {C.}~\bibnamefont {Hampel}}, \bibinfo
  {author} {\bibfnamefont {A.}~\bibnamefont {Hesselmann}}, \bibinfo {author}
  {\bibfnamefont {G.}~\bibnamefont {Hetzer}}, \bibinfo {author} {\bibfnamefont
  {T.}~\bibnamefont {Hrenar}}, \bibinfo {author} {\bibfnamefont
  {G.}~\bibnamefont {Jansen}}, \bibinfo {author} {\bibfnamefont
  {C.}~\bibnamefont {K{\"{o}}ppl}}, \bibinfo {author} {\bibfnamefont
  {Y.}~\bibnamefont {Liu}}, \bibinfo {author} {\bibfnamefont {A.~W.}\
  \bibnamefont {Lloyd}}, \bibinfo {author} {\bibfnamefont {R.~A.}\ \bibnamefont
  {Mata}}, \bibinfo {author} {\bibfnamefont {A.~J.}\ \bibnamefont {May}},
  \bibinfo {author} {\bibfnamefont {S.~J.}\ \bibnamefont {McNicholas}},
  \bibinfo {author} {\bibfnamefont {W.}~\bibnamefont {Meyer}}, \bibinfo
  {author} {\bibfnamefont {M.~E.}\ \bibnamefont {Mura}}, \bibinfo {author}
  {\bibfnamefont {A.}~\bibnamefont {Nicklass}}, \bibinfo {author}
  {\bibfnamefont {D.~P.}\ \bibnamefont {O'Neill}}, \bibinfo {author}
  {\bibfnamefont {P.}~\bibnamefont {Palmieri}}, \bibinfo {author}
  {\bibfnamefont {D.}~\bibnamefont {Peng}}, \bibinfo {author} {\bibfnamefont
  {K.}~\bibnamefont {Pfl{\"{u}}ger}}, \bibinfo {author} {\bibfnamefont
  {R.}~\bibnamefont {Pitzer}}, \bibinfo {author} {\bibfnamefont
  {M.}~\bibnamefont {Reiher}}, \bibinfo {author} {\bibfnamefont
  {T.}~\bibnamefont {Shiozaki}}, \bibinfo {author} {\bibfnamefont
  {H.}~\bibnamefont {Stoll}}, \bibinfo {author} {\bibfnamefont {A.~J.}\
  \bibnamefont {Stone}}, \bibinfo {author} {\bibfnamefont {R.}~\bibnamefont
  {Tarroni}}, \bibinfo {author} {\bibfnamefont {T.}~\bibnamefont
  {Thorsteinsson}}, \ and\ \bibinfo {author} {\bibfnamefont {M.}~\bibnamefont
  {Wang}},\ }\href@noop {} {\enquote {\bibinfo {title} {Molpro, version 2015.1,
  a package of ab initio programs},}\ }\bibinfo {howpublished}
  {http://www.molpro.net} (\bibinfo {year} {2015})\BibitemShut {NoStop}%
\bibitem [{\citenamefont {Dunning}(1989)}]{89Dunning.ai}%
  \BibitemOpen
  \bibfield  {author} {\bibinfo {author} {\bibfnamefont {T.~H.}\ \bibnamefont
  {Dunning}},\ }\href {\doibase 10.1063/1.456153} {\bibfield  {journal}
  {\bibinfo  {journal} {J. Chem. Phys.}\ }\textbf {\bibinfo {volume} {90}},\
  \bibinfo {pages} {1007} (\bibinfo {year} {1989})}\BibitemShut {NoStop}%
\bibitem [{\citenamefont {Balabanov}\ and\ \citenamefont
  {Peterson}(2005)}]{05BaPexx.ai}%
  \BibitemOpen
  \bibfield  {author} {\bibinfo {author} {\bibfnamefont {N.~B.}\ \bibnamefont
  {Balabanov}}\ and\ \bibinfo {author} {\bibfnamefont {K.~A.}\ \bibnamefont
  {Peterson}},\ }\href {\doibase 10.1063/1.1998907} {\bibfield  {journal}
  {\bibinfo  {journal} {J. Chem. Phys.}\ }\textbf {\bibinfo {volume} {123}},\
  \bibinfo {pages} {064107} (\bibinfo {year} {2005})}\BibitemShut {NoStop}%
\bibitem [{\citenamefont {Hopkins}, \citenamefont {Hamilton},\ and\
  \citenamefont {Mackenzie}(2009)}]{09HoHaMa.VO}%
  \BibitemOpen
  \bibfield  {author} {\bibinfo {author} {\bibfnamefont {W.~S.}\ \bibnamefont
  {Hopkins}}, \bibinfo {author} {\bibfnamefont {S.~M.}\ \bibnamefont
  {Hamilton}}, \ and\ \bibinfo {author} {\bibfnamefont {S.~R.}\ \bibnamefont
  {Mackenzie}},\ }\href {\doibase 10.1063/1.3104844} {\bibfield  {journal}
  {\bibinfo  {journal} {J. Chem. Phys.}\ }\textbf {\bibinfo {volume} {130}},\
  \bibinfo {pages} {144308} (\bibinfo {year} {2009})}\BibitemShut {NoStop}%
\bibitem [{\citenamefont {Neese}(2012)}]{12ORCA.programs}%
  \BibitemOpen
  \bibfield  {author} {\bibinfo {author} {\bibfnamefont {F.}~\bibnamefont
  {Neese}},\ }\href {\doibase 10.1002/wcms.81} {\bibfield  {journal} {\bibinfo
  {journal} {Wiley Interdiscip. Rev.-Comput. Mol. Sci.}\ }\textbf {\bibinfo
  {volume} {2}},\ \bibinfo {pages} {73} (\bibinfo {year} {2012})}\BibitemShut
  {NoStop}%
\bibitem [{\citenamefont {Aidas}\ \emph {et~al.}(2014)\citenamefont {Aidas},
  \citenamefont {Angeli}, \citenamefont {Bak}, \citenamefont {Bakken},
  \citenamefont {Bast}, \citenamefont {Boman}, \citenamefont {Christiansen},
  \citenamefont {Cimiraglia}, \citenamefont {Coriani}, \citenamefont {Dahle},
  \citenamefont {Dalskov}, \citenamefont {Ekstr{\"{o}}m}, \citenamefont
  {Enevoldsen}, \citenamefont {Eriksen}, \citenamefont {Ettenhuber},
  \citenamefont {Fern{\'{a}}ndez}, \citenamefont {Ferrighi}, \citenamefont
  {Fliegl}, \citenamefont {Frediani}, \citenamefont {Hald}, \citenamefont
  {Halkier}, \citenamefont {H{\"{a}}ttig}, \citenamefont {Heiberg},
  \citenamefont {Helgaker}, \citenamefont {Hennum}, \citenamefont {Hettema},
  \citenamefont {Hjertenaes}, \citenamefont {H{\o}st}, \citenamefont
  {H{\o}yvik}, \citenamefont {Iozzi}, \citenamefont {Jans{\'{i}}k},
  \citenamefont {Jensen}, \citenamefont {Jonsson}, \citenamefont
  {J{\o}rgensen}, \citenamefont {Kauczor}, \citenamefont {Kirpekar},
  \citenamefont {Kjaergaard}, \citenamefont {Klopper}, \citenamefont {Knecht},
  \citenamefont {Kobayashi}, \citenamefont {Koch}, \citenamefont {Kongsted},
  \citenamefont {Krapp}, \citenamefont {Kristensen}, \citenamefont {Ligabue},
  \citenamefont {Lutnaes}, \citenamefont {Melo}, \citenamefont {Mikkelsen},
  \citenamefont {Myhre}, \citenamefont {Neiss}, \citenamefont {Nielsen},
  \citenamefont {Norman}, \citenamefont {Olsen}, \citenamefont {Olsen},
  \citenamefont {Osted}, \citenamefont {Packer}, \citenamefont {Pawlowski},
  \citenamefont {Pedersen}, \citenamefont {Provasi}, \citenamefont {Reine},
  \citenamefont {Rinkevicius}, \citenamefont {Ruden}, \citenamefont {Ruud},
  \citenamefont {Rybkin}, \citenamefont {Sa{\l}ek}, \citenamefont {Samson},
  \citenamefont {de~Mer{\'{a}}s}, \citenamefont {Saue}, \citenamefont {Sauer},
  \citenamefont {Schimmelpfennig}, \citenamefont {Sneskov}, \citenamefont
  {Steindal}, \citenamefont {Sylvester-Hvid}, \citenamefont {Taylor},
  \citenamefont {Teale}, \citenamefont {Tellgren}, \citenamefont {Tew},
  \citenamefont {Thorvaldsen}, \citenamefont {Th{\o}gersen}, \citenamefont
  {Vahtras}, \citenamefont {Watson}, \citenamefont {Wilson}, \citenamefont
  {Ziolkowski},\ and\ \citenamefont {{\AA}gren}}]{14DALTON.method}%
  \BibitemOpen
  \bibfield  {author} {\bibinfo {author} {\bibfnamefont {K.}~\bibnamefont
  {Aidas}}, \bibinfo {author} {\bibfnamefont {C.}~\bibnamefont {Angeli}},
  \bibinfo {author} {\bibfnamefont {K.~L.}\ \bibnamefont {Bak}}, \bibinfo
  {author} {\bibfnamefont {V.}~\bibnamefont {Bakken}}, \bibinfo {author}
  {\bibfnamefont {R.}~\bibnamefont {Bast}}, \bibinfo {author} {\bibfnamefont
  {L.}~\bibnamefont {Boman}}, \bibinfo {author} {\bibfnamefont
  {O.}~\bibnamefont {Christiansen}}, \bibinfo {author} {\bibfnamefont
  {R.}~\bibnamefont {Cimiraglia}}, \bibinfo {author} {\bibfnamefont
  {S.}~\bibnamefont {Coriani}}, \bibinfo {author} {\bibfnamefont
  {P.}~\bibnamefont {Dahle}}, \bibinfo {author} {\bibfnamefont {E.~K.}\
  \bibnamefont {Dalskov}}, \bibinfo {author} {\bibfnamefont {U.}~\bibnamefont
  {Ekstr{\"{o}}m}}, \bibinfo {author} {\bibfnamefont {T.}~\bibnamefont
  {Enevoldsen}}, \bibinfo {author} {\bibfnamefont {J.~J.}\ \bibnamefont
  {Eriksen}}, \bibinfo {author} {\bibfnamefont {P.}~\bibnamefont {Ettenhuber}},
  \bibinfo {author} {\bibfnamefont {B.}~\bibnamefont {Fern{\'{a}}ndez}},
  \bibinfo {author} {\bibfnamefont {L.}~\bibnamefont {Ferrighi}}, \bibinfo
  {author} {\bibfnamefont {H.}~\bibnamefont {Fliegl}}, \bibinfo {author}
  {\bibfnamefont {L.}~\bibnamefont {Frediani}}, \bibinfo {author}
  {\bibfnamefont {K.}~\bibnamefont {Hald}}, \bibinfo {author} {\bibfnamefont
  {A.}~\bibnamefont {Halkier}}, \bibinfo {author} {\bibfnamefont
  {C.}~\bibnamefont {H{\"{a}}ttig}}, \bibinfo {author} {\bibfnamefont
  {H.}~\bibnamefont {Heiberg}}, \bibinfo {author} {\bibfnamefont
  {T.}~\bibnamefont {Helgaker}}, \bibinfo {author} {\bibfnamefont {A.~C.}\
  \bibnamefont {Hennum}}, \bibinfo {author} {\bibfnamefont {H.}~\bibnamefont
  {Hettema}}, \bibinfo {author} {\bibfnamefont {E.}~\bibnamefont {Hjertenaes}},
  \bibinfo {author} {\bibfnamefont {S.}~\bibnamefont {H{\o}st}}, \bibinfo
  {author} {\bibfnamefont {I.-M.}\ \bibnamefont {H{\o}yvik}}, \bibinfo {author}
  {\bibfnamefont {M.~F.}\ \bibnamefont {Iozzi}}, \bibinfo {author}
  {\bibfnamefont {B.}~\bibnamefont {Jans{\'{i}}k}}, \bibinfo {author}
  {\bibfnamefont {H.~J.~A.}\ \bibnamefont {Jensen}}, \bibinfo {author}
  {\bibfnamefont {D.}~\bibnamefont {Jonsson}}, \bibinfo {author} {\bibfnamefont
  {P.}~\bibnamefont {J{\o}rgensen}}, \bibinfo {author} {\bibfnamefont
  {J.}~\bibnamefont {Kauczor}}, \bibinfo {author} {\bibfnamefont
  {S.}~\bibnamefont {Kirpekar}}, \bibinfo {author} {\bibfnamefont
  {T.}~\bibnamefont {Kjaergaard}}, \bibinfo {author} {\bibfnamefont
  {W.}~\bibnamefont {Klopper}}, \bibinfo {author} {\bibfnamefont
  {S.}~\bibnamefont {Knecht}}, \bibinfo {author} {\bibfnamefont
  {R.}~\bibnamefont {Kobayashi}}, \bibinfo {author} {\bibfnamefont
  {H.}~\bibnamefont {Koch}}, \bibinfo {author} {\bibfnamefont {J.}~\bibnamefont
  {Kongsted}}, \bibinfo {author} {\bibfnamefont {A.}~\bibnamefont {Krapp}},
  \bibinfo {author} {\bibfnamefont {K.}~\bibnamefont {Kristensen}}, \bibinfo
  {author} {\bibfnamefont {A.}~\bibnamefont {Ligabue}}, \bibinfo {author}
  {\bibfnamefont {O.~B.}\ \bibnamefont {Lutnaes}}, \bibinfo {author}
  {\bibfnamefont {J.~I.}\ \bibnamefont {Melo}}, \bibinfo {author}
  {\bibfnamefont {K.~V.}\ \bibnamefont {Mikkelsen}}, \bibinfo {author}
  {\bibfnamefont {R.~H.}\ \bibnamefont {Myhre}}, \bibinfo {author}
  {\bibfnamefont {C.}~\bibnamefont {Neiss}}, \bibinfo {author} {\bibfnamefont
  {C.~B.}\ \bibnamefont {Nielsen}}, \bibinfo {author} {\bibfnamefont
  {P.}~\bibnamefont {Norman}}, \bibinfo {author} {\bibfnamefont
  {J.}~\bibnamefont {Olsen}}, \bibinfo {author} {\bibfnamefont {J.~M.~H.}\
  \bibnamefont {Olsen}}, \bibinfo {author} {\bibfnamefont {A.}~\bibnamefont
  {Osted}}, \bibinfo {author} {\bibfnamefont {M.~J.}\ \bibnamefont {Packer}},
  \bibinfo {author} {\bibfnamefont {F.}~\bibnamefont {Pawlowski}}, \bibinfo
  {author} {\bibfnamefont {T.~B.}\ \bibnamefont {Pedersen}}, \bibinfo {author}
  {\bibfnamefont {P.~F.}\ \bibnamefont {Provasi}}, \bibinfo {author}
  {\bibfnamefont {S.}~\bibnamefont {Reine}}, \bibinfo {author} {\bibfnamefont
  {Z.}~\bibnamefont {Rinkevicius}}, \bibinfo {author} {\bibfnamefont {T.~A.}\
  \bibnamefont {Ruden}}, \bibinfo {author} {\bibfnamefont {K.}~\bibnamefont
  {Ruud}}, \bibinfo {author} {\bibfnamefont {V.~V.}\ \bibnamefont {Rybkin}},
  \bibinfo {author} {\bibfnamefont {P.}~\bibnamefont {Sa{\l}ek}}, \bibinfo
  {author} {\bibfnamefont {C.~C.~M.}\ \bibnamefont {Samson}}, \bibinfo {author}
  {\bibfnamefont {A.~S.}\ \bibnamefont {de~Mer{\'{a}}s}}, \bibinfo {author}
  {\bibfnamefont {T.}~\bibnamefont {Saue}}, \bibinfo {author} {\bibfnamefont
  {S.~P.~A.}\ \bibnamefont {Sauer}}, \bibinfo {author} {\bibfnamefont
  {B.}~\bibnamefont {Schimmelpfennig}}, \bibinfo {author} {\bibfnamefont
  {K.}~\bibnamefont {Sneskov}}, \bibinfo {author} {\bibfnamefont {A.~H.}\
  \bibnamefont {Steindal}}, \bibinfo {author} {\bibfnamefont {K.~O.}\
  \bibnamefont {Sylvester-Hvid}}, \bibinfo {author} {\bibfnamefont {P.~R.}\
  \bibnamefont {Taylor}}, \bibinfo {author} {\bibfnamefont {A.~M.}\
  \bibnamefont {Teale}}, \bibinfo {author} {\bibfnamefont {E.~I.}\ \bibnamefont
  {Tellgren}}, \bibinfo {author} {\bibfnamefont {D.~P.}\ \bibnamefont {Tew}},
  \bibinfo {author} {\bibfnamefont {A.~J.}\ \bibnamefont {Thorvaldsen}},
  \bibinfo {author} {\bibfnamefont {L.}~\bibnamefont {Th{\o}gersen}}, \bibinfo
  {author} {\bibfnamefont {O.}~\bibnamefont {Vahtras}}, \bibinfo {author}
  {\bibfnamefont {M.~A.}\ \bibnamefont {Watson}}, \bibinfo {author}
  {\bibfnamefont {D.~J.~D.}\ \bibnamefont {Wilson}}, \bibinfo {author}
  {\bibfnamefont {M.}~\bibnamefont {Ziolkowski}}, \ and\ \bibinfo {author}
  {\bibfnamefont {H.}~\bibnamefont {{\AA}gren}},\ }\href {\doibase
  10.1002/wcms.1172} {\bibfield  {journal} {\bibinfo  {journal} {Wiley
  Interdiscip. Rev.-Comput. Mol. Sci.}\ }\textbf {\bibinfo {volume} {4}},\
  \bibinfo {pages} {269} (\bibinfo {year} {2014})}\BibitemShut {NoStop}%
\bibitem [{\citenamefont {Schweiger}\ and\ \citenamefont
  {Jeschke}(2001)}]{01ScJexx.epr}%
  \BibitemOpen
  \bibfield  {author} {\bibinfo {author} {\bibfnamefont {A.}~\bibnamefont
  {Schweiger}}\ and\ \bibinfo {author} {\bibfnamefont {G.}~\bibnamefont
  {Jeschke}},\ }\href@noop {} {\emph {\bibinfo {title} {Principles of pulse
  electron paramagnetic resonance}}}\ (\bibinfo  {publisher} {Oxford University
  Press},\ \bibinfo {year} {2001})\BibitemShut {NoStop}%
\bibitem [{\citenamefont {Slotterback}\ \emph {et~al.}(1995)\citenamefont
  {Slotterback}, \citenamefont {Clement}, \citenamefont {Janda},\ and\
  \citenamefont {Western}}]{95SlClJa.hyperfine}%
  \BibitemOpen
  \bibfield  {author} {\bibinfo {author} {\bibfnamefont {T.~J.}\ \bibnamefont
  {Slotterback}}, \bibinfo {author} {\bibfnamefont {S.~G.}\ \bibnamefont
  {Clement}}, \bibinfo {author} {\bibfnamefont {K.~C.}\ \bibnamefont {Janda}},
  \ and\ \bibinfo {author} {\bibfnamefont {C.~M.}\ \bibnamefont {Western}},\
  }\href {\doibase 10.1063/1.470023} {\bibfield  {journal} {\bibinfo  {journal}
  {J. Chem. Phys.}\ }\textbf {\bibinfo {volume} {103}},\ \bibinfo {pages}
  {9125} (\bibinfo {year} {1995})}\BibitemShut {NoStop}%
\bibitem [{\citenamefont {Tennyson}, \citenamefont {Hill},\ and\ \citenamefont
  {Yurchenko}(2013)}]{jt548}%
  \BibitemOpen
  \bibfield  {author} {\bibinfo {author} {\bibfnamefont {J.}~\bibnamefont
  {Tennyson}}, \bibinfo {author} {\bibfnamefont {C.}~\bibnamefont {Hill}}, \
  and\ \bibinfo {author} {\bibfnamefont {S.~N.}\ \bibnamefont {Yurchenko}},\
  }in\ \href {\doibase 10.1063/1.4815853} {\emph {\bibinfo {booktitle}
  {6$^{th}$ international conference on atomic and molecular data and their
  applications ICAMDATA-2012}}},\ \bibinfo {series} {AIP Conference
  Proceedings}, Vol.\ \bibinfo {volume} {1545}\ (\bibinfo  {publisher} {AIP,
  New York},\ \bibinfo {year} {2013})\ pp.\ \bibinfo {pages}
  {186--195}\BibitemShut {NoStop}%
\bibitem [{\citenamefont {Yurchenko}, \citenamefont {Al-Refaie},\ and\
  \citenamefont {Tennyson}(2018)}]{jt708}%
  \BibitemOpen
  \bibfield  {author} {\bibinfo {author} {\bibfnamefont {S.~N.}\ \bibnamefont
  {Yurchenko}}, \bibinfo {author} {\bibfnamefont {A.~F.}\ \bibnamefont
  {Al-Refaie}}, \ and\ \bibinfo {author} {\bibfnamefont {J.}~\bibnamefont
  {Tennyson}},\ }\href {\doibase 10.1051/0004-6361/201732531} {\bibfield
  {journal} {\bibinfo  {journal} {Astron. Astrophys.}\ }\textbf {\bibinfo
  {volume} {614}},\ \bibinfo {pages} {A131} (\bibinfo {year}
  {2018})}\BibitemShut {NoStop}%
\bibitem [{\citenamefont {Karlsson}\ \emph {et~al.}(1997)\citenamefont
  {Karlsson}, \citenamefont {Lindgren}, \citenamefont {Lundevall},\ and\
  \citenamefont {Sassenberg}}]{97KaLiLuSa}%
  \BibitemOpen
  \bibfield  {author} {\bibinfo {author} {\bibfnamefont {L.}~\bibnamefont
  {Karlsson}}, \bibinfo {author} {\bibfnamefont {B.}~\bibnamefont {Lindgren}},
  \bibinfo {author} {\bibfnamefont {C.}~\bibnamefont {Lundevall}}, \ and\
  \bibinfo {author} {\bibfnamefont {U.}~\bibnamefont {Sassenberg}},\ }\href
  {\doibase 10.1006/jmsp.1996.7173} {\bibfield  {journal} {\bibinfo  {journal}
  {J. Mol. Spectrosc.}\ }\textbf {\bibinfo {volume} {181}},\ \bibinfo {pages}
  {274} (\bibinfo {year} {1997})}\BibitemShut {NoStop}%
\end{thebibliography}%

\end{document}